\documentclass[twocolumn]{article}%
\usepackage{arxiv}
\usepackage[authoryear]{natbib}
\setcitestyle{aysep={}}
\usepackage[colorlinks,allcolors=blue]{hyperref}
\usepackage{amsthm}
\usepackage{etoolbox}
\makeatletter

\newcommand{\fnm}[1]{#1}
\newcommand{\sur}[1]{#1}
\newcommand{\orgname}[1]{#1}
\newcommand{\orgaddress}[1]{#1}
\newcommand{\city}[1]{#1}
\newcommand{\country}[1]{#1}
\newcommand{\orcid}[1]{\,\href{https://orcid.org/#1}{\includegraphics[height=1.6ex]{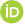}}}
\newif\ifarx@firstauthor \arx@firstauthortrue
\renewcommand{\author}{\@ifstar{\arx@author}{\arx@author}}
\newcommand{\arx@author}[2][]{%
  \ifarx@firstauthor\gdef\@author{}\global\arx@firstauthorfalse\else\g@addto@macro\@author{\And}\fi
  \g@addto@macro\@author{#2\textsuperscript{#1}}}
\newcommand{\email}[1]{\g@addto@macro\@author{\\\normalfont\texttt{#1}}}
\gdef\@date{}
\newcommand{\affil}[2][]{\g@addto@macro\@date{\textsuperscript{#1}#2\par}}
\let\arx@abstractbegin\abstract
\let\arx@abstractend\endabstract
\renewcommand{\abstract}[1]{\gdef\arx@abstracttext{#1}}
\let\arx@keywords\keywords
\renewcommand{\keywords}[1]{\gdef\arx@keywordstext{#1}}
\let\arx@maketitle\maketitle
\renewcommand{\maketitle}{\twocolumn[{\arx@maketitle
  \arx@abstractbegin\arx@abstracttext\arx@abstractend
  \arx@keywords{\arx@keywordstext}\vskip 2em}]}
\newcommand{\bmhead}[1]{\subsubsection*{#1}}
\providecommand{\backmatter}{}

\renewcommand{\shorttitle}{CodeHealth Predicts the Token Cost of Agentic Code Review}
\renewcommand{\headeright}{}
\renewcommand{\undertitle}{}

\def\arx@tabletype{table}
\newif\ifarx@intable
\g@addto@macro\@floatboxreset{\ifx\@captype\arx@tabletype\centering\arx@intabletrue\fi}
\AtBeginEnvironment{tabular}{\arx@intablefalse}
\AfterEndEnvironment{tabular}{\ifarx@intable\par\vskip\smallskipamount
  \leftskip\z@skip\rightskip\z@skip\parfillskip\z@ plus 1fil\relax\fi}

\renewenvironment{table}
  {\setlength{\abovecaptionskip}{\@belowcaptionskip}%
   \setlength{\belowcaptionskip}{\@abovecaptionskip}%
   \@dblfloat{table}}
  {\end@dblfloat}
\renewenvironment{figure}{\@dblfloat{figure}}{\end@dblfloat}

\AtBeginDocument{\DeclareRobustCommand{\[}{\arx@display}}
\def\arx@display#1\]{\begin{equation*}\sbox\z@{$\displaystyle#1$}%
  \ifdim\wd\z@>\displaywidth\resizebox{\displaywidth}{!}{\usebox\z@}\else\usebox\z@\fi
  \end{equation*}}
\usepackage{xurl}

\makeatother

\usepackage{graphicx}%
\usepackage{multirow}%
\usepackage{tabularx}%
\usepackage{array}%
\usepackage{amsmath,amssymb,amsfonts}%
\usepackage{amsthm}%
\usepackage{mathrsfs}%
\usepackage[title]{appendix}%
\usepackage{xcolor}%
\usepackage{textcomp}%
\usepackage{manyfoot}%
\usepackage{booktabs}%
\usepackage{algorithm}%
\usepackage{algorithmicx}%
\usepackage{algpseudocode}%
\usepackage{listings}%
\usepackage{caption}%
\usepackage{subcaption}%
\definecolor{shadecolor}{gray}{0.9}%
\usepackage{framed}%
\usepackage{breakurl}
\usepackage[english]{babel} %
\usepackage{csquotes}%
\usepackage{tikz}%
\usetikzlibrary{arrows.meta,positioning,fit,backgrounds,calc,shapes.geometric,matrix}%

\newcommand{\Section}[1]{Section~\ref{#1}}
\newcommand{\Figure}[1]{Fig.~\ref{#1}}
\newcommand{\Table}[1]{Table~\ref{#1}}

\newcommand{\RQ}[1]{\texorpdfstring{\textbf{RQ\textsubscript{#1}}}{RQ#1}}
\newcommand{\RQanswer}[2]{\begin{shaded}\noindent\textbf{#1:} #2\end{shaded}}

\newcolumntype{L}{>{\raggedright\arraybackslash}X}

\begin{document}

\title{CodeHealth Predicts the Token Cost of Agentic Code Review: A Pre-Registered Study of 465 Agent-Authored Pull Requests}

\author*[1]{\fnm{Lo} \sur{Gullstrand Heander} \orcid{0000-0002-0695-4580}}\email{lo.gullstrand\_heander@cs.lth.se}

\author[1,2]{\fnm{Markus} \sur{Borg} \orcid{0000-0001-7879-4371}}\email{markus.borg@cs.lth.se}

\author[1]{\fnm{Emma} \sur{S\"oderberg} \orcid{0000-0001-7966-4560}}\email{emma.soderberg@cs.lth.se}

\author[1]{\fnm{Christofer} \sur{Rydenf\"alt} \orcid{0000-0003-1495-8263}}\email{christofer.rydenfalt@design.lth.se}

\affil[1]{\orgname{Lund University}, \orgaddress{\city{Lund}, \country{Sweden}}}
\affil[2]{\orgname{CodeScene AB}, \orgaddress{\city{Malmö}, \country{Sweden}}}

\abstract{
\textbf{Context:} Organizations increasingly delegate code review to multi-agent large language model (LLM) pipelines billed per token. Low code quality raises human maintenance effort, but its effect on agent review cost has not been measured.

\textbf{Objective:} We test whether baseline CodeHealth predicts the token cost of agentic review and fix beyond pull request (PR) footprint.

\textbf{Method:} In a pre-registered observational study, we sampled 465 agent-authored PRs from 246 repositories in AIDev, jointly stratified by CodeHealth and footprint under repository caps. A frozen five-agent review-and-fix pipeline processed every PR with three different LLMs. Per LLM, we fitted a mixed model of log tokens on CodeHealth and log PR footprint, with a repository random intercept.

\textbf{Results:} For all three LLMs the CodeHealth coefficient was negative, with 95\% CI excluding zero; each CodeHealth point was associated with 3.5\% to 6.1\% lower token cost. CodeHealth was unrelated to PR footprint, so PR size does not explain the association. Two of three registered smells predicted fix cost; batch fixing left median CodeHealth unchanged. Exploratory analyses located the association in the review stage, where agents iterated more on unhealthy code without writing more comments. Adjusting for the size of the scored files absorbed the association, which we interpret as mediation by file size.

\textbf{Conclusions:} A quality metric calibrated against human maintainability also predicts the inference cost of multi-agent review and fix. The effect is modest per point and not carried by PR size, so the refactoring case extends to agent-reviewed code when it changes structure and file size, not only the score.

\textbf{Data Availability:} \href{https://doi.org/10.5281/zenodo.22127920}{DOI:10.5281/zenodo.22127920}
}


\maketitle

\section{Introduction}
\label{sec:introduction}

Technical debt research has linked low code quality to higher maintenance costs for over 30 years. \citet{cunningham1992wycash} coined the term when he described shipping not-quite-right code as going into technical debt on which interest is paid in every subsequent change, and over time this metaphor has turned into a research field with its own systematic maps and syntheses~\citep{li_systematic_2015, ampatzoglou_financial_2015}. Researchers have also quantified the cost; for example, \citet{besker_software_2019}, replicating and extending an earlier study of developers' own working time, found that developers waste on average 23\% of their time on technical debt. Across this literature, the unit of cost is human: attention, working time, and the effort of rebuilding an understanding that the code itself fails to convey.

That unit of cost is changing as organizations route code review through multi-agent large language model (LLM) pipelines to absorb rising pull request (PR) volumes. \citet{he_writes_2026} report that in one enterprise where AI tool adoption roughly doubled per-engineer PR throughput, automated review overtook human code review, while the load on each remaining human reviewer roughly doubled. When the reviewer is an agent, its cost stops being time, attention, and effort and instead becomes metered token cost, which has not yet been related to the quality of the code under review. \citet{storey_debt_2026} argues that generative AI does not remove debt so much as redistribute it, separating technical debt in the code from cognitive debt in a team's shared understanding and intent debt in its externalized rationale. She observes that agents working against poor context may \enquote{use more tokens and time than expected}. The observation is plausible and, to our knowledge, untested. If it is right, the long-standing case for refactoring extends into workflows with no human reviewer in them; if it is wrong, teams that budget refactoring and maintenance on the strength of human-effort evidence have to justify that spend on other grounds for automated pipelines.

CodeHealth\texttrademark{} (CH) is a composite, deterministic score between $1$ and $10$ that aggregates several source-level quality signals into a single file-level measure. Beyond its links to human maintenance effort, it has been shown to predict how AI tooling behaves on the code: \citet{borg_code_2026} report significantly lower break rates when AI refactors healthy code. This study asks whether the same metric predicts the token cost of multi-agent review and fix of a PR. The question cannot be asked without separating quality from size. Token consumption scales mechanically with the size of a diff, which presents more input tokens, more lines to read, and more edits to make, so if unhealthy code tends to arrive in larger PRs, an apparent relationship between CH and token cost may simply reflect PR footprint.

To study this relationship, we conducted a pre-registered observational study investigating whether baseline CodeHealth predicts agentic review and fix costs beyond what PR footprint explains. We treat PR footprint as a mediator candidate, not a covariate, and we designed the study under equipoise so that either outcome is informative. If baseline CH predicts the cost of agentic review and fix after controlling for footprint, the case for refactoring extends to agentic review workflows. If the effect of CH on token cost is instead fully mediated by PR size, the economic argument shifts toward keeping PRs small, possibly at some expense of code quality.

We investigate four pre-registered\footnote{OSF DOI: \href{https://doi.org/10.17605/OSF.IO/FBJCT}{10.17605/OSF.IO/FBJCT}} research questions:

\begin{description}
  \item[\RQ{1}] Does baseline CodeHealth predict the token cost of agentic review and fix, controlling for PR footprint, language, and repository?
  \item[\RQ{2}] Is any such relationship direct, or is it mediated by the size of the PR that unhealthy code attracts?
  \item[\RQ{3}] Do specific code smells, rather than the aggregate CodeHealth score, predict the cost of the fix stage in particular?
  \item[\RQ{4}] What happens to CodeHealth itself when a batch fix agent applies a full set of review comments in a single pass?
\end{description}

\begin{figure}[tbh]
\centering
\begin{subfigure}[t]{0.49\linewidth}
\includegraphics[width=\linewidth]{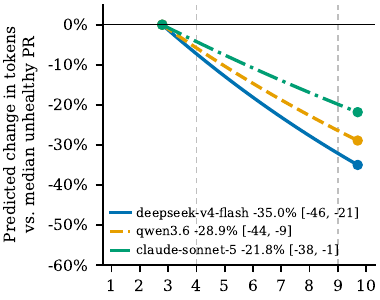}
\caption{\RQ{1}: Predicted change in $\mathit{tokens\_total}$ from the median unhealthy-tier PR (CH 2.8) to the median healthy-tier PR (CH 9.7). The legend gives the total change across the 6.9-point contrast with its 95\% confidence interval. Dashed verticals mark the tier boundaries at CH 4 and CH 9.}
\label{fig:intro-rq1}
\end{subfigure}\hfill
\begin{subfigure}[t]{0.49\linewidth}
\includegraphics[width=\linewidth]{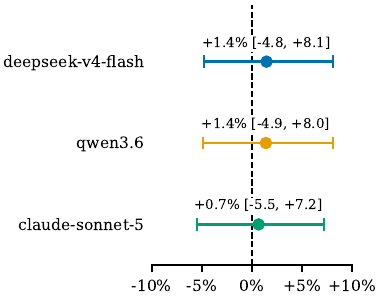}
\caption{\RQ{2}: Each LLM's registered CH-to-footprint correlation as percent change in changed lines per CH point with its 95\% confidence interval.}
\label{fig:intro-rq2}
\end{subfigure}
\caption{The registered answers to \RQ{1} and \RQ{2}, fitted separately per LLM. \textbf{Takeaway:} Token cost falls with CH for all three LLMs, and CH does not predict footprint, so larger PR footprints do not account for the extra cost of unhealthy code.}
\label{fig:intro-results}
\end{figure}

The registered analysis supports the direct-CodeHealth-effect branch of this equipoise. Fitted separately for each of the three LLMs on the achieved 465-PR sample, the coefficient on baseline CH is negative, with a 95\% confidence interval excluding zero, for all three LLMs. This satisfies \RQ{1}'s pre-registered conjunctive criterion. One additional CH point is associated with an estimated 3.5\% to 6.1\% reduction in total review-and-fix tokens, holding footprint, language, and repository constant (\Figure{fig:intro-rq1}). Footprint itself is essentially unrelated to baseline CodeHealth in this sample (at most \(+0.014\) in \(\log\) footprint per CH point, with every interval spanning zero), so the effect is direct with respect to PR footprint and not an artifact of unhealthy code attracting larger PRs (\Figure{fig:intro-rq2}). A quality metric calibrated against human maintainability therefore also shows an association with metered agent inference cost. Even if this is not established as a causal relationship, the result suggests that the code properties the metric penalizes also raise what an agent spends. In an exploratory analysis, the association concentrates in the review stage: agents reviewing unhealthy code take more turns and consume more context without raising more comments. A second exploratory refit shows that the association cannot be separated from the size of the files the agents read, which we interpret as file size, unlike PR footprint, mediating the effect (\Section{sec:discussion-filesize}). Keeping PRs small therefore does not substitute for code quality, and the case for refactoring extends to agent-reviewed code in the form that changes the structure and size of the files an agent reads through, not only the score (\Section{sec:discussion-roads-to-health}).

In addressing these questions, we make four contributions:
\begin{enumerate}
  \item A pre-registered, multi-LLM mixed-effects observational study across three independently served LLMs, evaluated with a conjunctive direction-consistency criterion.
  \item A frozen, reproducible review-and-fix harness of four review agents and one fix agent grounded in the ISO/IEC 25010 product quality model~\citep{iso_25010_2023}, with all agent prompts hashed and registered, and unchanged across the whole collection.
  \item A jointly stratified CodeHealth-by-footprint, repository-capped sample of 465 agent-authored PRs drawn from 246 distinct repositories in the AIDev dataset~\citep{li_aidev_2026}, accompanied by an audit of its achieved balance.
  \item A documented account of two sampling failure modes specific to AIDev and one pooling artifact in multi-LLM PR studies, together with the decomposition checklist that detected them (\Section{sec:experiment-design} and \Section{sec:threats-conclusion}).
\end{enumerate}

\section{Background}
\label{sec:background}

Code review has been studied since Fagan introduced formal inspections as a defect-removal technique~\citep{fagan_design_1976}, with tool support since ICICLE offered network-distributed inspections in 1990~\citep{brothers_icicle_1990}. \citet{bacchelli_expectations_2013} showed that its lighter, tool-based contemporary form, \textbf{modern code review}, also delivers benefits beyond defect finding, such as knowledge transfer, team awareness, and alternative solutions to problems. Work in industrial~\citep{sadowski_modern_2018, macleod_code_2018} and open-source settings~\citep{mcintosh_empirical_2016, kononenko_code_2016, baysal_investigating_2016} studied the practice in situ and documented its variability~\citep{felderer_choice_2017} and challenges~\citep{soderberg_understanding_2022}, with both its benefits and its costs measured in human units.

Cost has been a persistent theme in this literature, although most often measured indirectly. \citet{goncalves_explicit_2022} framed review as an expensive process and studied whether explicit review strategies could improve effectiveness and efficiency. \citet{kudrjavets_mining_2022} quantified the delay between acceptance and merge as a component of a team's code velocity. \citet{dorner_upper_2025} estimated the upper bound of information diffusion achievable through review as a communication network. \citet{yang_roadmap_2026}, in a recent roadmap, described manual inspection of increasingly complex systems as cognitively demanding and resource-intensive, and identified the resulting workflow bottleneck as a primary motivation for automation. Across all of these, the unit of cost is human: attention, time, or effort. Our own prior work takes the same human-centered stance, modeling review as a decision-making process driven by the questions reviewers ask~\citep{gullstrand_heander_code_2026} and arguing for AI that supports rather than replaces the reviewer~\citep{heander_support_2025}, a position shared by the Copenhagen Manifesto on human-centered generative AI in software engineering~\citep{russo_generative_2024}.

\subsection{CodeHealth as a Measurable Construct}
\label{sec:bg-health}

\textbf{CodeHealth\texttrademark{}} (CH) is an aggregate, file-level static-analysis score in the range 1 to 10, calculated by CodeScene\footnote{\url{https://codescene.com/}} from a catalog of source-level code smells such as complex methods, deeply nested conditionals, and duplicated logic~\citep{codescene_health}. CodeScene divides this scale into three levels, called red, yellow, and green code on its product pages~\citep{codescene_health} and unhealthy, problematic, and healthy in its scale definition~\citep{codescene_scores}: below 4, from 4 up to 9, and at 9 and above. We adopt the same thresholds and use the second set of names, \emph{unhealthy}, \emph{problematic}, and \emph{healthy}, throughout this paper. Unlike a single structural metric, CH is designed as a measurable proxy for maintainability and has been evaluated against organizational outcomes. \citet{tornhill_code_2022} analyzed 39 proprietary production codebases and reported that files with low CH scores took considerably longer in development and contained more defects than files with higher scores. \citet{borg_increasing_2024} subsequently found that the benefit of raising CH holds even when the score is already high, that is, the returns do not flatten out at the top of the scale. \citet{borg_owns_2023} found that low CH slows issue resolution, particularly for developers with marginal ownership of the code. CodeHealth has also been benchmarked against human judgment: \citet{borg_ghost_2024} compared maintainability metrics and machine-learning predictions against human assessments and reported where the automated measures agree and diverge (\Section{sec:threats-construct}).

Recent work connects CodeHealth to AI-assisted development. In a two-phase pre-registered controlled experiment methodologically close to the present study, \citet{borg_echoes_2025} found no significant difference in CH when manually evolving code that was either co-developed with AI or originally human-only. \citet{borg_code_2026} reported that low CH predicts a higher risk that AI coding tools introduce new defects, and reframed the metric as a potential predictor of AI-friendliness as well as of human maintainability. Most recently, \citet{wirdemann_code_2026} applied the same reframing to a different task, single-shot LLM-generated unit tests over a competitive-programming corpus. Their study reported both that CH correlates with generated-test effectiveness and, more relevant here, that source code files with lower CH require more input tokens, an association that persisted after controlling for source lines of code in Java and C++ but largely disappeared in the smaller, healthier Python subset. Outside CodeScene, \citet{ramirez_exploring_2025} related a different composite, the Maintainability Index, to agent issue resolution on SWE-bench Lite and found correlations that varied across LLMs. That the association survives a change of task, corpus, and outcome variable suggests that CodeHealth may predict, more generally, how much work an LLM needs to reason about code. If so, it should also predict how many tokens an LLM spends reviewing and repairing that code. This inference is plausible but untested, and it is the premise of the present study. Our registered model adjusts for the number of changed lines, not for the size of the files the agents read; we report a size-adjusted refit as an exploratory analysis (\Section{sec:results-filesize}).

\subsection{Agentic Code Review}
\label{sec:bg-agentic}

Automated review has moved from single-model comment generation~\citep{tufano_using_2022} to multi-agent architectures in which specialized agents examine a PR from different angles and a supervisory component aggregates their findings, as in CodeAgent~\citep{tang_codeagent_2024}, ChatDev~\citep{qian_chatdev_2024}, and AutoCodeRover~\citep{zhang_autocoderover_2024}, which \citet{he_llm-based_2025} describe as a pivotal evolution in LLM-based software engineering. Deployed systems follow the same pattern, from BitsAI-CR, which pairs a rule-based detector with an LLM filter stage and reports 75\% precision at ByteDance~\citep{sun_bitsai_2025}, to the system of \citet{laiq_using_2026} at Ericsson, in which four agents each cover one quality dimension and developers rated 96\% of 206 generated findings correct and 69\% of those important enough to fix.

What this literature predominantly reports is quality and adoption: precision of generated comments, F1 on comment classification, developer-rated correctness and importance~\citep{laiq_using_2026}, developer adoption rate, and PR cycle time. Inference cost as a function of the properties of the code under review is not among the reported outcomes. The closest work we are aware of is the PR triage proposal of \citet{madeyski_triage_2026}, which suggests using CodeHealth as a routing signal for assigning software engineering tasks to the cheapest sufficient LLM tier, based on a hypothesized tier-dependent asymmetry in which mid-capability LLMs benefit from clean code while frontier (flagship-tier) LLMs do not. The routing design uses SWE-bench Lite as an evaluation benchmark and concerns task solving instead of review. It presupposes a measured association between CH and inference cost on real PRs, which has not yet been reported.

\subsection{CodeHealth and Agentic Review Cost}
\label{sec:bg-gap}

To our knowledge, no study has related CodeHealth to the cost of automated code review. In one enterprise that adopted AI coding tools, automated review overtook human review while per-reviewer load roughly doubled~\citep{he_writes_2026}, yet what that automation costs, and what in the code drives the cost, is not reported. Teams that justify refactoring budgets by human maintenance effort therefore have no evidence on whether the same code quality also governs the metered cost of the agents now reviewing their code. CH has been validated as a predictor of human maintenance effort~\citep{tornhill_code_2022, borg_increasing_2024} and of AI defect-injection risk and generated-test effectiveness~\citep{borg_code_2026, wirdemann_code_2026}, but not of the cost of automated code review, and the agentic review studies above report precision and adoption, but not cost. This study estimates that relationship.
Answering the question requires separating CodeHealth from PR footprint, since token cost is mechanically coupled to diff size, yet footprint cannot simply be adjusted away: if unhealthy code forces larger PRs for comparable tasks, adjusting for size would erase part of the effect of interest. \Section{sec:method} therefore specifies footprint as a mediator candidate and reports the decomposition.

\section{Method}
\label{sec:method}

\begin{figure}[hb!t]
\centering
\includegraphics[width=0.8\textwidth]{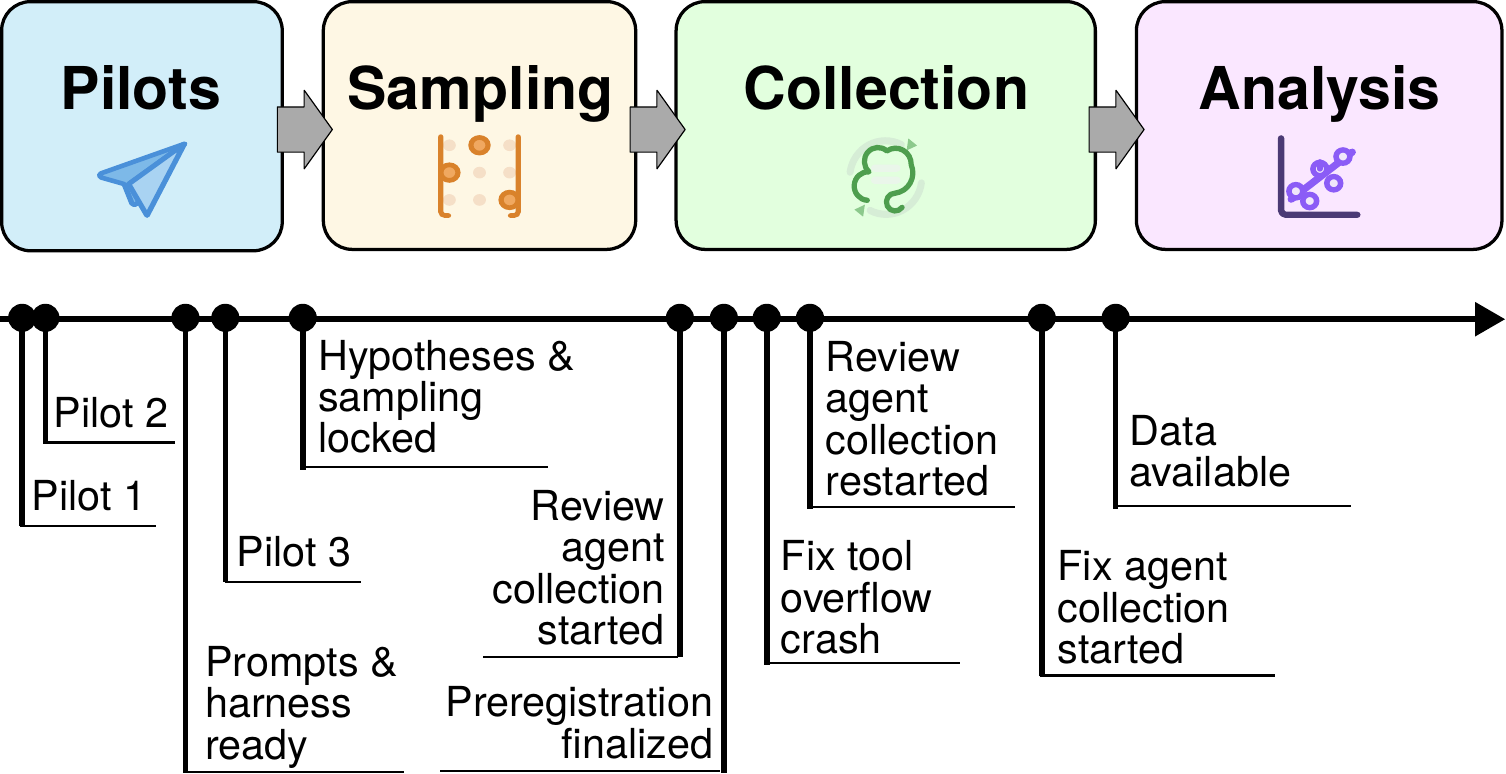}
\caption{Overview of the study procedure in four phases, with its milestones on a timeline. \textbf{Takeaway:} Hypotheses and sampling were locked after pilot 3, and the pre-registration was finalized after the review agents had started but before the fix agent ran, so no complete row existed at filing (\Section{sec:method-prereg}).}
\label{fig:overview}
\end{figure}

This is a non-randomized, observational study of a fixed measurement instrument applied to a stratified sample of PRs. It was pre-registered 2026-08-20 on the Open Science Framework (OSF; DOI: \href{https://doi.org/10.17605/OSF.IO/FBJCT}{10.17605/OSF.IO/FBJCT}), and the protocol described here is the registered protocol. \Figure{fig:overview} summarizes the four phases of the procedure (pilots, sampling, collection, and analysis), together with the milestones within them. The pre-registration falls inside the collection phase, after the review agents had started and before the fix agent ran, with the chronology given in \Section{sec:method-prereg}.

\subsection{Design and Hypotheses}
\label{sec:method-design}

The unit of analysis is the PR. For each PR, a fixed pipeline of four review agents and one fix agent is run against the repository with the PR's changes applied, and the token cost of that run is recorded. Nothing about the pipeline is manipulated and no condition is randomized; the pipeline is a measurement instrument held constant across all 465 observations. What varies between observations is the code the instrument is applied to.

The predictor of interest is the code's baseline CodeHealth, aggregated over the files the PR touches and weighted by the number of lines the PR changes in each of them (\Section{sec:method-measures}). Four properties of this choice govern the interpretation of the resulting estimate. First, the measurement is taken on the PR's head commit, so it scores exactly the code the review agents are pointed at, and it is taken before any agent of ours has run, so it cannot be an effect of the pipeline. Second, and consequently, it includes the PR's own contribution instead of measuring only the code the PR landed on: the analysis tool scores whole files, not line ranges, so on a PR that edits established files the score comes largely from pre-existing content, while on a PR that introduces new files it is a property of the submitted patch itself. In our sample, the former case dominates and the distribution is quantified in \Section{sec:threats-construct}. We call the predictor \emph{baseline} in the sense of before review, not before the PR. Third, it is scoped to the changed files instead of the whole repository, so the predictor characterizes the files in which the review and fix work happens and excludes code that is unaffected. Fourth, the line weighting means that baseline CH and PR footprint are two measurements over the same set of files, correlated by construction, and separating them is the subject of \RQ{2}. Taken together, we treat CH as a property of the code a review agent is handed, which a maintainer influences both through the standing state of the repository and through what is allowed into it, and we ask how many tokens one point of CodeHealth is associated with.

Because CodeHealth is directly observed, the estimate is associational, not causal. Unhealthy code is not distributed at random across repositories, languages, or problem domains. The design absorbs two confounding factors, repository identity as a random intercept and primary language as a fixed factor, leaving the remainder as a threat described in more detail in \Section{sec:threats-internal}. The same 465 PRs are processed independently by three different LLMs, producing three correlated replications of the same estimate rather than one estimate on three times as much data, an important distinction for the analysis plan in \Section{sec:method-analysis}.

The quantity estimated by the study is \(\beta_{\mathit{health}}\), the coefficient on baseline CodeHealth in the primary model below. The outcome ($\mathit{tokens\_total}$) enters that model as a natural logarithm while the predictor enters untransformed, so \(\beta_{\mathit{health}}\) is the expected change in \(\log\) total tokens associated with a one-point increase in baseline CH, holding footprint, language, and repository fixed. Exponentiating it, \(100(e^{\beta_{\mathit{health}}} - 1)\), re-expresses the same quantity as a percentage change in predicted token cost per CH point, which is the form a practitioner weighing a refactoring investment can act on. A negative \(\beta_{\mathit{health}}\) therefore means that healthier code is cheaper for an agent to review and repair at a given PR size, and its magnitude is the effect of one CH point measured in tokens. Because CodeHealth is bounded on a 1 to 10 scale, the coefficient also scales to a contrast between tiers. In the achieved sample, the median unhealthy-tier PR scores 2.8 and the median healthy-tier PR scores 9.7, so \(6.9\beta_{\mathit{health}}\) is the difference between a typical unhealthy and a typical healthy PR. Throughout, we interpret \(\beta_{\mathit{health}}\) as an association within the fitted model, not as a causal effect.

The confirmatory hypothesis, registered verbatim, is the following.

\begin{shaded}
  \textbf{H1}: In \(\log(\mathit{tokens\_total}) \sim \mathit{health} + \log(\mathit{footprint}) + \mathit{language} + (1|\mathit{repo})\), fitted separately for each of the three LLMs on the same 465 PRs, the coefficient on baseline CodeHealth is negative.
\end{shaded}

\begin{figure}[thb]
\centering
\begin{tikzpicture}[
  font=\small,
  var/.style={draw, rounded corners=2pt, align=center, inner sep=5pt, minimum height=9mm, fill=white},
  obs/.style={var, thick},
  path/.style={-{Latex[length=2.4mm]}, thick},
  weak/.style={-{Latex[length=2.4mm]}, thick, densely dashed, gray!70!black},
  plab/.style={font=\small, inner sep=2pt, fill=white}
]
  \node[obs] (health) at (0,0) {Baseline CodeHealth\\{\footnotesize changed files, at PR head}};
  \node[var] (fp) at (4.6,2.0) {$\log$ change footprint\\{\footnotesize changed lines}};
  \node[obs] (cost) at (9.2,0) {$\log$ \textit{tokens\_total}\\{\footnotesize review $+$ fix}};
  \node[var] (lang) at (4.6,-1.9) {Language\\{\footnotesize fixed factor, 5 levels}};
  \node[inner sep=0pt] (headroom) at (4.6,2.95) {};

  \draw[path] (health) -- node[plab, above left=-1pt and 0pt] {$a$ (\RQ{2})} (fp);
  \draw[path] (fp) -- node[plab, above right=-1pt and 0pt] {$b$ (\RQ{2})} (cost);
  \draw[path] (health) -- node[plab, below] {$c'$ \;\; \textbf{H1}: $c' < 0$ (\RQ{1})} (cost);
  \draw[weak] (lang) -- (cost);

  \begin{scope}[on background layer]
    \node[draw, densely dotted, thick, rounded corners=4pt, inner sep=10pt, fill=black!3,
          fit=(health)(fp)(cost)(lang)(headroom)] (plate) {};
  \end{scope}
  \node[anchor=north west, font=\footnotesize, align=left] at ([xshift=4pt, yshift=-4pt] plate.north west)
    {465 PRs nested in 246 repositories: $(1|\mathit{repo})$ random intercept};
  \node[anchor=north east, font=\footnotesize, align=right] at ([xshift=-2pt, yshift=-3pt] plate.south east)
    {Fitted separately for each of the three LLMs: no pooling, no CodeHealth $\times$ LLM interaction.};
\end{tikzpicture}
\vspace{2mm}
\caption{Associational path diagram of the registered model, \(\log(\mathit{tokens\_total}) \sim \mathit{health} + \log(\mathit{footprint}) + \mathit{language} + (1|\mathit{repo})\). \textbf{Takeaway:} Change footprint is treated as a mediator candidate sitting on the path from CodeHealth to cost, not as a nuisance covariate to be removed, so \RQ{1} concerns the direct path \(c'\) that survives conditioning on footprint and \RQ{2} concerns the decomposition of \(a\), \(b\), and \(c'\); the arrows denote regression paths in an observational design and do not imply causality.}
\label{fig:pathmodel}
\end{figure}
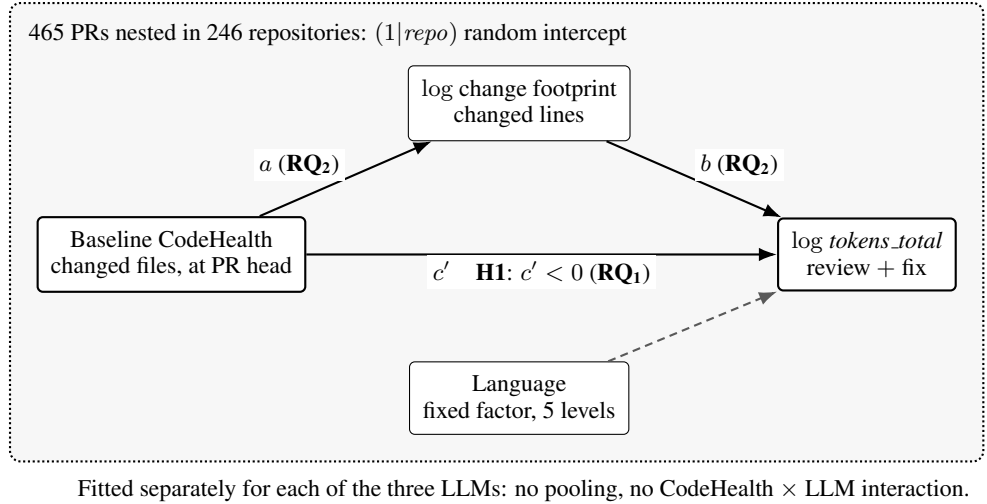

\Figure{fig:pathmodel} shows the variables of the primary model and the roles they play in it. The success criterion was pre-stated and is conjunctive. H1 is supported if and only if all three per-LLM estimates of \(\beta_{\mathit{health}}\) are negative \emph{and} at least two of the three 95\% confidence intervals exclude zero. If all three estimates are negative but at most one confidence interval excludes zero, the result is reported as direction consistent but individually underpowered, and no confirmatory claim is made. If the estimates disagree in sign, H1 is not supported and the study reports cost as footprint-mediated. \RQ{2} through \RQ{4} are registered as a mediation decomposition, a three-hypothesis smell family, and a descriptive distribution respectively (\Section{sec:method-analysis}). 

\subsection{Dataset}
\label{sec:method-dataset}

We draw PRs from AIDev, a dataset of PRs authored by AI coding agents on GitHub~\citep{li_aidev_2026}. AIDev provides a full tier of 932,791 PRs and a curated tier restricted to repositories with at least 100 GitHub stars, comprising 33,596 PRs across 2,807 repositories. We use the curated tier, because it is the only tier that includes the diffs and commit details needed to reconstruct a reviewable PR. The five originating agents recorded in the dataset are OpenAI Codex, Devin, GitHub Copilot, Cursor, and Claude Code. The dataset is revision-locked and SHA-256-validated at download time (revision \texttt{68ed5f4b}, SHA-256 \texttt{ec75e016}).

AIDev is large enough to support a jointly stratified design, and its curated tier removes the abandoned and single-author repositories that would otherwise populate the unhealthy stratum with code that nobody maintains and whose review cost nobody pays. Its PRs were authored by contemporary coding agents, not humans, which matches the setting the study is about; agentic review is increasingly applied to agent-written code, and \citet{he_writes_2026} report that in repositories adopting these agents, automated review has already overtaken human review. For the same reason, a repository that routinely merges agent-authored PRs has already made the configuration and process decisions that agent-driven development requires, so the sample is drawn from projects in which the practice is already established. The corresponding limitation, that the findings speak to agent-authored PRs in relatively popular open-source projects and not to human-authored, private, or legacy code, is stated in \Section{sec:threats-external}.

Two filters are applied in order: PR state must be \texttt{merged}, so that a well-defined base commit exists; and the repository's primary language must be one of the five most represented in the curated subset, namely TypeScript, Python, Go, C\#, and JavaScript, fixed in advance to avoid language-specific bias. The resulting candidate pool contains 19,403 PRs from 1,164 distinct repositories. A third requirement, that the PR has file-level patches available, is checked for each PR when it is measured during sampling, and a PR without them is never sampled. The pool is concentrated: a single repository, \texttt{mochilang/mochi}, supplies 39.3\% of it, the two largest repositories together supply 57.5\%, and 474 repositories (41\%) contribute exactly one PR. This concentration is the motivation for the sampling design below.

\subsection{Pilot Studies and Design Evolution}
\label{sec:experiment-design}

The design described in this section is the fourth data collection of the study. Here, we describe the three pilot exercises that preceded it, because the sampling design of \Section{sec:method-sampling}, the random intercept for repository, and several reporting conventions are direct responses to the ways in which the first two pilots failed. \Table{tab:pilots} summarizes all four collections. The pilots used disjoint PRs, all three are disclosed in the registration, and all 103 of their PR identifiers are permanently excluded from the main sample.

\begin{table}[t]
\caption{The three pilot exercises and the main collection. \textbf{Takeaway:} Pilots 1 and 2 failed for sampling reasons rather than pipeline reasons, and each failure influenced the final design; pilot 3 is the main collection at one tenth of the scale, run with the same three LLMs, prompts, and workflow, with small harness updates in between.}
\label{tab:pilots}
\begin{tabularx}{\textwidth}{@{}lr>{\raggedright\arraybackslash}p{2.1cm}LL@{}}
\toprule
\textbf{Collection} & \textbf{PRs} & \textbf{Date} & \textbf{Sampling and LLMs} & \textbf{Outcome} \\
\midrule \vspace{6pt}
Pilot 1 & 21 & 2026-06-26 & CodeHealth tier only, no repository cap; earlier prompts and harness; \texttt{qwen3-6} without prefix caching & Discarded: CodeHealth tier aliased with PR size \\ \vspace{6pt}
Pilot 2 & 37 & 2026-07-07 & CodeHealth tier only, no repository cap; \texttt{qwen3-6} without prefix caching & Discarded: unhealthy tier 92\% one repository \\ \vspace{6pt}
Pilot 3 & 45 & 2026-07-13 to 2026-08-19 & \(3 \times 3\) grid, 5 per cell, at most 1 PR per repository per cell and 3 overall; all three LLMs & Hypothesis-generating basis for the registration \\
Main & 465 & 2026-08-19 to 2026-08-24 & \(3 \times 3\) grid, streamed to exhaustion, at most 2 per cell and 9 overall; all three LLMs & Confirmatory \\
\bottomrule
\end{tabularx}
\end{table}

Pilots 1 and 2 were run under an earlier design of this study, which asked whether agentic review and fix degrades CodeHealth, and which compared the batch fix pass used here against a second, CodeHealth-guided fix condition. Both pilots ran only the batch fix condition, with a single locally served LLM (\texttt{qwen3-6}, at that time on a serving stack without prefix caching), and both drew PRs stratified on CH tier alone, in candidate order and without any cap on how many PRs one repository could contribute. Both produced results that did not survive the decomposition checklist, and we believe the two failure modes could generalize to other studies that sample from agent-authored PR datasets. Pilot 1 reported a strong negative correlation between baseline CH and reviewer tokens (\(r = -.68\), \(p < .001\)), but CH tier and PR size were fully aliased in that sample, since its healthy tier consisted of small PRs with a median of about 40 changed lines, so the CH variable was effectively a size variable. Pilot 2 broke the aliasing on the healthy side and the correlation vanished, but its unhealthy tier was 12 of 13 PRs (92\%) from the single repository \texttt{mochilang/mochi}, so it measured the properties of one codebase and not of CodeHealth in general. That concentration was close to inevitable, since the same repository supplies 39.3\% of the candidate pool (\Section{sec:method-dataset}). Neither the apparent effect of pilot 1 nor the apparent null of pilot 2 therefore generalizes, and both are retained only as diagnostic exhibits.

Pilot 2 also changed the question. In 26 of its 37 PRs (70\%), CodeHealth after the batch fix was exactly equal to CH before it, so a study of degradation under batch fix would have had almost no effective sample, and the CodeHealth-guided condition tested a treatment that conflates the change under review with unrelated refactoring. We therefore dropped the second condition, turned the batch fix pass into the fixed instrument of \Section{sec:method-instrument}, made cost the primary outcome, and retained \(\Delta\)CodeHealth as the descriptive \RQ{4}. The two sampling failures produced four further changes: joint stratification on CH and footprint, to break the aliasing of pilot 1; hard repository caps, to break the concentration of pilot 2 (\Section{sec:method-sampling}); a random intercept for repository in the primary model (\Section{sec:method-analysis}); and a standing \emph{decomposition checklist} applied before any aggregate is trusted. The checklist consists of a comparison of means against medians, a footprint-adjusted cost per changed line, an outlier check for whether any single PR accounts for more than 20\% of a stratum's sum, a rank correlation and a Mann-Whitney contrast on the continuous predictor, and an audit of the repository-by-cell matrix. We tested it against the false results of pilots 1 and 2 and verified that it raises warnings in both cases.

Pilot 3 implemented the final study design at a smaller scale. It drew five PRs per cell of the same \(3 \times 3\) grid, 45 in total, under tight repository caps of one PR per repository per cell and three overall, which yielded 38 distinct repositories and a baseline CodeHealth that is orthogonal to footprint in the sample (\(\rho = .015\)). It ran the full instrument with all three LLMs of the main collection: \texttt{deepseek-v4-flash} and \texttt{sonnet-5} on 2026-07-13 and 2026-07-14, and \texttt{qwen3-6}, re-run after its move to the final serving stack of \Section{sec:method-models}, on 2026-08-18 and 2026-08-19. Pilot 3 is the disclosed, hypothesis-generating basis of the registration. It supplied the expected direction of H1 and the sample-size rationale (\Section{sec:method-prereg}), the three confirmatory code smells of \RQ{3}, and the decision to fit the model separately per LLM instead of pooling the LLMs' data (\Section{sec:method-analysis}). It is not counted as confirmatory evidence.

The main collection repeats the setup of pilot 3, with two differences. The first is scale. Sampling streamed the entire candidate pool instead of stopping at five PRs per cell, and the repository caps were relaxed to the values in \Section{sec:method-sampling}, because the tight caps of pilot 3 left mostly single-PR repositories and therefore little within-repository variation from which the random intercept could separate repository from CodeHealth. The second is the harness. The main collection uses the same agent prompts, LLMs, serving stack, and review and fix workflow as pilot 3. To avoid a failure mode with spiraling token costs that could inflate the effect size, character caps on tool output were implemented on 2026-08-21 and filed as an amendment to the registration (\Section{sec:method-prereg}).

\subsection{Sampling}
\label{sec:method-sampling}

We sample on a \(3 \times 3\) grid formed by crossing three CodeHealth tiers with three PR-footprint terciles. The CH tiers, introduced in \Section{sec:bg-health}, are the strata published by the tool vendor~\citep{codescene_health}: \emph{unhealthy} below 4, \emph{problematic} from 4 up to 9, and \emph{healthy} at 9 and above. The footprint terciles are log-scale cut points on changed lines, computed from the entire filtered candidate pool and reused unchanged for every sampling run. \emph{Small} PRs have 59 changed lines or fewer, \emph{medium} PRs 60--177 lines, and \emph{large} PRs 178 lines or more.

Two independent repository caps are enforced during sampling: at most two PRs from any one repository within a single cell, and at most nine PRs from any one repository across the whole sample. These caps are a design requirement because the primary model estimates the CodeHealth effect partly from within-repository contrasts, which it cannot do if a single repository dominates a stratum. Sampling proceeds under a fixed seed with per-PR checkpoints, so that interrupting and re-running reproduces the same sample. We streamed the entire candidate pool to exhaustion and collected the largest dataset possible under these constraints. All 103 PRs previously included in any of the pilot runs were excluded.

\subsection{Achieved Sample}
\label{sec:method-sample}

The achieved sample contains 465 PRs from 246 distinct repositories, at a mean of 1.89 PRs per repository. \Table{tab:cellfill} shows the fill per cell against the nominal per-cell target of 56.

\begin{table}[t]
\caption{Achieved cell fill on the \(3 \times 3\) CodeHealth by footprint grid, against a target of 56 PRs per cell. \textbf{Takeaway:} Seven of nine cells filled completely. The two short cells are in the unhealthy tier at medium and large footprint, reflecting a real scarcity of low-CH PRs in the AIDev pool.}
\label{tab:cellfill}
\begin{tabular}{lrrrr}
\toprule
\textbf{CodeHealth tier} & \textbf{Small PRs} & \textbf{Medium PRs} & \textbf{Large PRs} & \textbf{Tier total} \\
\midrule
Unhealthy    & 56 & 33 & 40 & \textbf{129} \\
Problematic  & 56 & 56 & 56 & \textbf{168} \\
Healthy      & 56 & 56 & 56 & \textbf{168} \\
\midrule
Tercile total & 168 & 145 & 152 & \textbf{465} \\
\bottomrule
\end{tabular}
\end{table}

As \Table{tab:cellfill} shows, only \emph{unhealthy-medium} (33 of 56) and \emph{unhealthy-large} (40 of 56) fell short. The sampler streamed all 19,403 candidates and the short cells are caused by low-CodeHealth PRs being scarce among merged PRs in repositories with at least 100 stars, and scarcer still as diff size grows. The unhealthy tier, not the total PR count, is the limiting constraint on sample size.

As designed, no repository contributes more than two PRs to any single cell. Four repositories reach the overall cap of nine, and 160 repositories (65\%) contribute exactly one PR. Of the 246 repositories, 55 (22\%) span at least two CodeHealth tiers, and these are the repositories from which the random intercept model estimates the within-repository CH effect. Baseline CH has a median of 8.14 and a mean of 6.95; changed lines have a median of 93 and a mean of 832, with a maximum of 91,128. The heavy-tailed shape is the rationale for reporting medians over means in \Section{sec:method-analysis}. The language distribution is TypeScript 174, Python 134, C\# 71, Go 51, and JavaScript 35. The originating coding agent is OpenAI Codex for 229 PRs (49\%), GitHub Copilot for 103 (22\%), Devin for 75 (16\%), Cursor for 43 (9\%), and Claude Code for 15 (3\%); sampling was not stratified on implementation language or originating agent, so this distribution is descriptive only.

\subsection{The Measurement Instrument}
\label{sec:method-instrument}

\begin{figure}[thb]
\centering
\includegraphics[width=\textwidth]{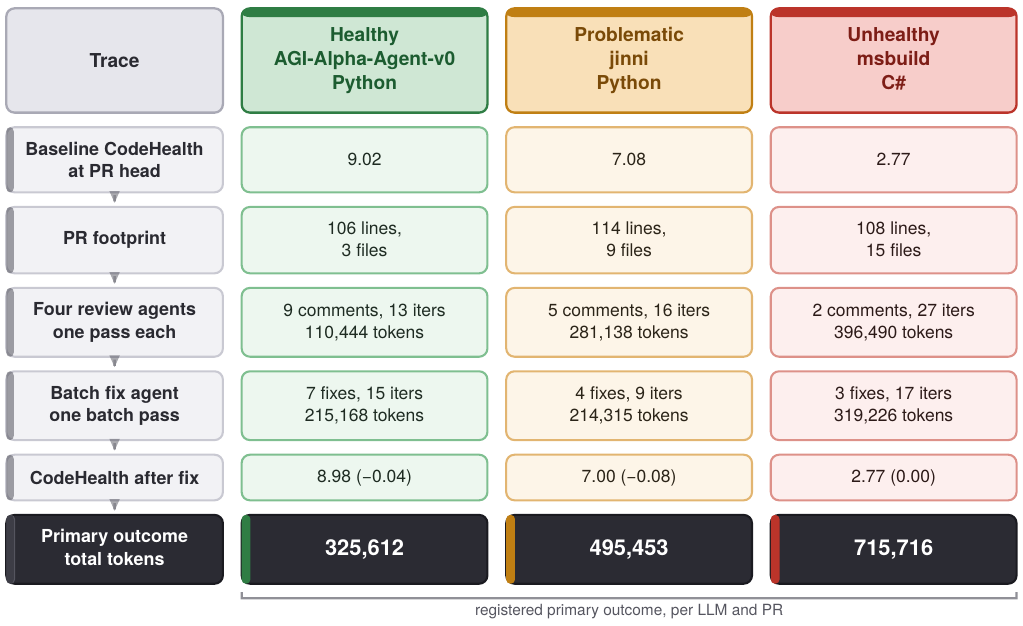}
\caption{One PR from each CodeHealth tier traced through the measurement instrument, all three from the medium footprint tercile and all three measured with \texttt{sonnet-5}. A pass is one invocation of an agent over its whole input; the iteration counts in each cell are the agent turns taken inside that single pass, so a batch fix pass with no retry can still run many iterations. Each is the PR whose total token cost lies closest to its own tier's median among PRs with 95 to 125 changed lines, so none is an extreme case. \textbf{Takeaway:} At a nearly identical PR footprint, total cost rises as baseline CH falls, and the rise is not explained by the number of review comments, which are unrelated to CH across the whole sample (\(\rho = -.03\), \(p = .56\)). Agent iterations, in contrast, are correlated with CH (\(\rho = -.16\), \(p < .001\)). Three PRs illustrate the measurement path and are not evidence for it; the estimates are in \Section{sec:results}.}
\label{fig:threepr}
\end{figure}

The instrument is a fixed four-agent review pass followed by a single batch fix pass, run once per PR per LLM. \Figure{fig:threepr} traces three concrete PRs, one from each CodeHealth tier, through its stages. For each PR, the repository is cloned into a shared bare-clone cache and a dedicated \texttt{git worktree} is created at the PR's head commit with a detached HEAD, which isolates concurrent PR runs from one another and presents the agents with the repository in the state a human reviewer would open it in, with the PR applied. Baseline CH is measured on that worktree, over the PR's changed files, before any agent runs. Each of the four review agents then examines the diff in a single pass and emits structured comments, each with a file, a line, an issue description, a severity, and an impact statement. One fix agent receives all comments from all four reviewers, constructs fixes for the issues, and applies them in a single batch pass. CH is measured a second time on the resulting worktree, and the worktree is removed.

Each agent runs a plain multi-turn tool loop. The LLM is called with a system prompt, the diff, and a set of tool definitions; any tool calls it emits are executed against the PR worktree and the results appended to the message history, until the LLM returns a final answer or a guard terminates it. The tool set exposes the working tree at the PR's head commit for reading, searching, and editing, and mirrors the read and search tools against the base commit, so that an agent can ask what a file, a hunk, or a match looked like before the PR without the worktree ever leaving the head state. The file-reading tools return at most 500 lines per call, prefixed with the file's total line count, so an agent pages through a long file in several calls. The editing tools are available only to the fix agent, so that a review agent cannot modify any file. A single adapter layer absorbs provider differences, so the same loop, the same tools, and the same prompts run against every LLM. Every LLM call in every agent loop is issued at temperature 0. Prompting is zero-shot throughout; no examples are included in any agent prompt.

\begin{table}[t]
\caption{The four review agents and their primary ISO/IEC 25010 quality characteristics~\citep{iso_25010_2023}. \textbf{Takeaway:} The agent set partitions review concerns along an externally defined taxonomy, which makes the choice of roles auditable; the mapping is approximate only for DesignReviewer, which spans two characteristics.}
\label{tab:agents}
\begin{tabularx}{\textwidth}{@{}lLL@{}}
\toprule
\textbf{Agent} & \textbf{Review scope} & \textbf{ISO/IEC 25010} \\
\midrule \vspace{6pt}
SecurityReviewer & Injection, authentication, cryptography, data
  exposure, path traversal, XSS, and CSRF & Security \\ \vspace{6pt}
PerformanceReviewer & Algorithmic complexity, memory, I/O, N+1 queries,
  and blocking asynchronous code & Performance Efficiency \\ \vspace{6pt}
DesignReviewer & Coupling and cohesion, layer violations, naming
  consistency, API design, and error propagation & Maintainability, Compatibility \\
LogicReviewer & State transitions, calculations, domain invariants,
  boundary errors, and control flow & Functional Suitability \\
\bottomrule
\end{tabularx}
\end{table}

\Table{tab:agents} lists the four roles, partitioned along the ISO/IEC 25010 taxonomy~\citep{iso_25010_2023}. Every review agent prompt restricts findings to issues introduced or exposed by the PR under review and excludes unchanged code from the review scope, so the review is bounded to the diff even though the tools give each agent access to the whole worktree at both the head and the base commit. The design excluded two scopes that other review sets include: a dedicated test reviewer and an intent-alignment reviewer. A dedicated test reviewer predominantly produces coverage-gap findings on files outside the diff, which have no baseline CodeHealth score and are therefore inert with respect to the measurement window. Intent alignment, meaning whether the patch implements what the PR description states, cannot be acted on by a fix agent without specifying the missing functionality; PR intent is instead passed to the fix agent as context, so that its edits do not contradict the rationale of the PR.

Each agent prompt was hashed with SHA-256 and the hashes were filed in the OSF pre-registration. The prompts, the orchestration logic, and the LLM identities are frozen; any change after the filing requires an OSF amendment, even for defects discovered during execution. One such defect was found during the main collection, in the tool layer, and \Section{sec:method-prereg} reports the required OSF amendment.

\subsection{LLMs and Infrastructure}
\label{sec:method-models}

\begin{table}[thb]
\caption{The three LLMs, their serving stacks, versions, and cached input shares. The cached input share is computed as the share of all input tokens served from the prefix cache across the 465-PR collection, pooled over review and fix agents. \textbf{Takeaway:} The three instruments differ simultaneously in provider, quantization, serving stack, and prefix-cache regime, which is why the analysis fits them separately and never compares them inferentially.}
\label{tab:models}
\begin{tabular}{llll}
\toprule
Identifier & Stack & Version & Cached input \\
\midrule
\texttt{deepseek/deepseek-v4-flash} & Hosted API & Provider, v0731 & $93.2\%$ \\
\texttt{nvidia/Qwen3.6-35B-A3B-NVFP4} & Local vLLM & Commit \texttt{491c2f1}, & $81.9\%$ \\
                                      &            & \texttt{vllm==0.27.1} & \\
\texttt{anthropic/claude-sonnet-5} & Hosted API & Provider, v5 & $89.7\%$ \\
\bottomrule
\end{tabular}
\end{table}

Three LLMs, listed in \Table{tab:models}, were used in the study. They were chosen to span deployment regimes plausible in real agentic code review: one inexpensive hosted API LLM, one open-weight LLM served locally on our own hardware, and one current-generation hosted commercial LLM.

The locally served LLM, \texttt{qwen3-6}, is pinned to a specific revision and its serving environment is pinned with a hash-checked lock file covering all 196 packages. Reasoning is enabled with a reasoning-token budget of 4,000 tokens and a maximum context length of 262,144 tokens. The main collection was executed between 2026-08-19 and 2026-08-24. Review sessions for two of the LLMs began one day before the registration was filed; \Section{sec:method-prereg} gives the chronology. The two hosted LLMs are versioned by their providers, so their pinning is provider-side and we cannot verify weight-level stability across the collection window.

Prefix caching counts repeated context across agent-loop iterations differently for the three LLMs (\Table{tab:models}, rightmost column). This means that total tokens are not comparable \emph{across} LLMs. Our own pilot data shows the size of the effect. When the local LLM moved from a serving stack with no prefix caching to one where 80.6\% of its input tokens were served from cache, its footprint-to-cost rank correlation on an identical set of 45 pilot PRs fell from \(\rho = .50\) to \(\rho = .27\), with no change to the LLM, the prompts, or the sample. We therefore register no cross-LLM inferential comparison of token totals, pre-specify two cache-neutral secondary outcomes for the review stage (review iterations, and uncached input plus output tokens), and note that the within-LLM slopes H1 tests are unaffected, because each LLM's cache regime is constant across its own 465 PRs.

\subsection{Measures}
\label{sec:method-measures}

\paragraph{Predictor.} Baseline CodeHealth is the mean of the per-file CH scores over the files the PR changes, taken at the PR's head commit before the review-and-fix pipeline runs and weighted by the number of lines the PR changes in each file. It is measured using CodeScene's Model Context Protocol (MCP) tool locked to version \texttt{1.4.0}. Because of the weighting, a file with two changed lines does not move a PR's score as much as a file with two hundred. The three tiers are used as a stratifier for building the cell matrix, while the analysis model uses the continuous score. CH is a deterministic static analysis metric, so the relevant stability risk is version drift only. This is mitigated by locking the version for the duration of the experiment.

\paragraph{Primary outcome.} Total tokens per PR, defined as the sum of input and output tokens across the four review agents and the fix agent. Cached tokens are a reported subset of input tokens in this accounting, not an additional pool, and are recorded separately for cost weighting.

\paragraph{Secondary outcomes.} Review iterations, fix iterations, comment count per reviewer role, and \(\Delta\)CodeHealth, defined as the post-fix score minus the baseline score. Where the post-fix measurement returns no file scores, \(\Delta\)CodeHealth is recorded as missing, not as zero, and excluded pairwise, since the scale starts at 1.0 and an invalid score can only indicate measurement failure.

\paragraph{Covariates and mediator candidate.} We treat PR footprint ($\log$ changed lines) as a mediator candidate rather than a covariate. Repository language is treated as a fixed factor, and repository identity as a random intercept.

\subsection{Data Analysis}
\label{sec:method-analysis}

The primary model is a linear mixed model with a random intercept for repository,

\[
\log(\mathit{tokens\_total}) \sim \mathit{health} + \log(\mathit{footprint}) +
\mathit{language} + (1|\mathit{repo}),
\]

fitted with restricted maximum likelihood in statsmodels 0.14.6~\citep{seabold_statsmodels_2010} under Python 3.12, with language entered as a fixed factor using the most frequent language as reference, and two-sided \(\alpha = .05\) throughout. No random slopes are fitted, because 246 repositories at a mean of 1.89 PRs each cannot support them.

The model is fitted \emph{separately} for each of the three LLMs. The three fits share the same 465 PRs, and a PR that is intrinsically expensive to review is expensive for every LLM that reviews it: on pilot data, the within-PR residual intraclass correlation was .581, a design effect of 2.16, meaning pooled rows provide less than half their apparent information. Naive pooling would have turned a non-significant CodeHealth effect into an apparently significant one (\Section{sec:threats-conclusion}). A CH by LLM interaction term was rejected on interpretability grounds: the three LLMs differ simultaneously in provider, quantization, serving stack, cache regime, and loop behavior, so a significant interaction term would establish only that three heterogeneously served instruments behaved differently. The conjunctive success criterion for H1 follows from this design. Requiring that all three estimates agree in sign and that at least two confidence intervals exclude zero is a replication criterion, so no multiplicity correction is applied to it.

\RQ{2} is answered by a three-path decomposition on the same model family: the association from CodeHealth to \(\log\) footprint, the association from \(\log\) footprint to cost, and the direct CH-to-cost path conditional on footprint, which is the coefficient in the primary model. The decomposition is associational; because tasks are not matched across CH levels, the analysis cannot establish whether unhealthy code \emph{causes} larger PRs.

\RQ{3} tests three code smell categories, selected from pilot data by a deterministic rule that was fixed before the main collection ran: retain a smell only if it is present in at least 10 of the 45 pilot PRs, only if all three LLMs agree on the sign of its partial correlation after controlling for \(\log\) changed lines, and only if the mean adjusted \(|\rho|\) across the three LLMs is at least .20; then rank the survivors and take the top three. The full list of code smells with their correlations on the pilot PRs is available in the pre-registration package. \Table{tab:smells} shows the result. Each smell was registered together with a one-sentence mechanism. For Complex Method, tangled control flow forces the fix agent to trace several execution paths before it can safely edit any one of them. For Code Duplication, a fix must be located and reconciled across several call sites. For Complex Conditional, the agent must reason through each branch's interaction with the PR. \RQ{3} targets the fix stage because that is where the agent edits code, and prior work found that AI edits break unhealthy code more often~\citep{borg_code_2026}; complex and duplicated code were expected to require edits in more places or more reasoning per edit, while the review stage is covered by \RQ{1}. Each smell is tested as a partial Spearman correlation against \(\log_{1p}\) fix-agent tokens, controlling for \(\log\) changed lines, and Holm-corrected within this three-hypothesis family.

\begin{table}[t]
\caption{The three code smells registered as confirmatory hypotheses for \RQ{3}, selected from pilot data by the pre-specified rule. \textbf{Takeaway:} The rule retained the three smells present in enough pilot PRs, with a sign agreed by all three LLMs and the largest mean adjusted correlations; each was then registered with a mechanism of its own.}
\label{tab:smells}
\begin{tabular}{lrrl}
\toprule
Smell & PRs present & Mean adjusted \(|\rho|\) & Rank \\
\midrule
Complex Method       & 35 & .339 & 1 \\
Code Duplication     & 12 & .333 & 2 \\
Complex Conditional  & 20 & .311 & 3 \\
\bottomrule
\end{tabular}
\end{table}

\RQ{4} reports \(\Delta\)CodeHealth as a median and a distribution plot per LLM. No hypothesis test is pre-registered for it; it is registered as a description to be confirmed, not as an established result.

Cost and footprint are log-transformed, using \(\log_{1p}\) wherever a value can be exactly zero. Cost distributions are summarized with medians and box or violin plots rather than tier means, because a single outlier PR accounted for approximately 45\% of a tier's mean token total in pilot data. Every analysis not named above is declared exploratory wherever it appears and corrected under Benjamini-Hochberg.

\subsection{Attrition and Exclusion}
\label{sec:method-attrition}

\begin{table}[htb]
\caption{Agent-session termination categories and their treatment. \textbf{Takeaway:} A guard-stopped review session is kept because it is instrument behavior and part of the cost being measured; a guard-stopped fix session leaves no completed fix record or post-fix score, so that PR is excluded for that LLM. Infrastructure failures and unexpected errors are retried until success.}
\label{tab:termination}
\begin{tabularx}{\textwidth}{@{} l L L @{}}
\toprule
\textbf{Category} & \textbf{Termination reasons} & \textbf{Treatment} \\
\midrule
Complete       & \texttt{completed} & Kept, primary data \\
Guard-stopped, review & \texttt{ceiling}, \texttt{explore\_stall}, \texttt{write\_drought}, \texttt{spiral}, \texttt{truncated} & Kept, flagged for sensitivity check \\
 Guard-stopped, fix & same reasons & Excluded: no completed fix record \\
Infrastructure & \texttt{rate\_limited}, \texttt{server\_error} & Retried, never skipped \\
Error          & \texttt{parse\_error}, \texttt{unexpected\_error} & Retried, never skipped \\
\bottomrule
\end{tabularx}
\end{table}

Exclusion criteria were specified in advance. A PR is excluded from a given LLM's fit only if its repository is no longer clonable or its base or head commit no longer resolvable, if the analysis tool cannot score any changed file, or if no completed fix record exists for that LLM after the documented retry path. An excluded PR is reported as attrition for its cell and LLM and is not replaced. In total, 28 runs out of 1{,}395 are excluded, an average attrition of $2\%$. Every one of the 28 is a fix session that a loop guard stopped before it completed, 12 for \texttt{deepseek-v4-flash}, 4 for \texttt{qwen3-6}, and 12 for \texttt{sonnet-5}; none is an infrastructure failure.

\Table{tab:termination} shows how non-complete agent sessions are treated. Guard-stopped review sessions, in which a loop guard terminated a review agent after partial output, are kept: an agent that spirals on a difficult diff has incurred real cost, and filtering that cost out would understate what reviewing that diff actually takes. Pilot data motivated this choice at registration. In pilot 3, reviewer sessions terminated by truncation were statistically indistinguishable from the rest of the sample on baseline CodeHealth (Mann-Whitney \(p = .61\)), and excluding them \emph{strengthened} the CH slope for that LLM, from \(\beta = -0.089\) to \(\beta = -0.110\), so exclusion would have favored H1. The main collection does not reproduce that independence, as \Section{sec:results-sensitivity} reports. As registered, the primary fit is repeated for every LLM without the PRs that contain a non-complete agent session. In the retained main-collection sample, 103 \texttt{qwen3-6} PRs contain at least one guard-stopped review session, 120 sessions in total; the corresponding count is zero for the other two LLMs, whose sensitivity fits are identical to their primary fits.

Analysis is independent for each LLM. Because the three fits are separate by construction, a PR excluded for one LLM is retained for the other two.

\subsection{Replication Subsample for Instrument Stability}
\label{sec:method-replication}

The pipeline is run once per PR per LLM, so the main collection cannot separate the part of a PR's measured cost that is a stable property of the diff from run-to-run variation in the agents themselves. Temperature 0 does not make a served LLM deterministic, and an agentic tool loop amplifies whatever variation reaches it, because a single divergent tool call redirects the remainder of the trajectory and therefore the token count.

To measure that variation, we draw 27 PRs from the 465-PR sample, three per cell of the \(3 \times 3\) grid, under a fixed seed and a cap of one PR per repository, and run the full pipeline on them twice more with every LLM. Together with the main collection this gives three runs per PR per LLM. The instrument is unchanged across replicates. Selection uses predictors only and never conditions on an observed outcome: drawing the subsample from PRs already seen to spiral or truncate would estimate the variance of the tail rather than of the instrument. The subsample spans 27 distinct repositories, all five languages, all five originating coding agents, the full range of baseline CodeHealth, and footprints from 1 to 5,474 changed lines.

The replicates are analyzed by fitting an intercept-only mixed model with a random intercept per PR, separately for each LLM, which partitions the residual into a between-PR component and a run-to-run component. We report their ratio, \(\lambda\), and the implied reliability of a single run, ICC(\(k=1\)), in \Section{sec:results-lambda}. This analysis is exploratory: it is not part of the registration, and the replicates are not averaged into the confirmatory estimates, which remain based on a single run per PR per LLM as registered.

\subsection{Pre-Registration and Prior Evidence}
\label{sec:method-prereg}

The registration was made after the predictors had been measured and audited for balance. To use our hardware execution window, we started the review agents on 2026-08-19, while the registration was being written, and froze the registration on 2026-08-20 at 14:14. By then, \texttt{deepseek-v4-flash} had completed review sessions for 241 PRs and \texttt{qwen3-6} for 84, and these sessions remain in the analyzed corpus. Their token, iteration, and comment counts were stored on our cluster before the filing but were never read. The fix stage, which every analyzable row requires, started only after the filing: the earliest fix session in the collection began on 2026-08-20 at 18:23. No post-fix CodeHealth score, and therefore no complete row for any PR, existed when the registration was filed. The agent prompts were last changed on 2026-07-11, so every session, before and after the filing, ran the prompts whose hashes the registration records. The no-peeking constraint therefore depends on procedure for the review-stage sessions that predate the filing, and on construction for everything else. The filed registration stated that the pipeline had not yet been run on the sample; we corrected that statement in a transparency note filed on 2026-09-24 as an update to the registration, which records the chronology above and changes no hypothesis, design decision, or analysis. OSF keeps each update as a version of the same record, so the original filing, the amendment described below, and the note are all reachable through the registration DOI (\href{https://doi.org/10.17605/OSF.IO/FBJCT}{DOI:10.17605/OSF.IO/FBJCT}).

The registration's expected direction and sample-size rationale come from pilot 3 (\Section{sec:experiment-design}), which is disclosed in the filing as hypothesis-generating and not counted as confirmatory evidence. \Table{tab:pilot} reports its per-LLM estimates and the power characteristics that follow from them.

\begin{table}[t]
\caption{Pilot 3 per-LLM CodeHealth-to-cost slope and the resulting power characteristics at the achieved main-collection design ($n = 465$, 246 repositories). \textbf{Takeaway:} All three pilot slopes are negative and similar in magnitude, but every LLM's own required sample size exceeds the achieved 465, which is why no single LLM is designated primary and why the confirmatory claim is a direction-consistency criterion across all three.}
\label{tab:pilot}
\begin{tabular}{lrrlrrr}
\toprule
LLM & \(\beta_{\mathit{health}}\) & SE & 95\% CI & ICC & MDE\textsuperscript{a} & Required \(n\) \\
\midrule
\texttt{deepseek-v4-flash} & $-0.035$ & 0.048 & $[-0.129, 0.059]$ & .497 & 0.048 & 1{,}516 \\
\texttt{qwen3-6}           & $-0.068$ & 0.064 & $[-0.193, 0.058]$ & .714 & 0.067 & \phantom{0}861 \\
\texttt{sonnet-5}          & $-0.071$ & 0.072 & $[-0.213, 0.070]$ & .358 & 0.070 & \phantom{0}711 \\
\bottomrule
\end{tabular}
\footnotesize{\textsuperscript{a}Minimum detectable effect at the achieved design, in the same log-token units as \(\beta_{\mathit{health}}\), for a two-sided test at \(\alpha = .05\) with 80\% power and a Kish design-effect correction for repository clustering.}
\end{table}

\Table{tab:pilot} illustrates a limitation of this study. All three pilot slopes are negative and similar in magnitude, with rank correlations in a narrow band from \(\rho = -.13\) to \(\rho = -.20\), but none reached significance at \(n = 45\). Additionally, every LLM's estimate of required sample size exceeds the 465 PRs possible to extract from the AIDev dataset under the stricter stratification and repository representation rules. At the selected design, the minimum detectable effect per CH point is between 0.048 and 0.070 in log-token units, which is a reduction of between 4.7\% and 6.7\% per point, enough only for two out of three pilot point estimates. For this reason, H1 is registered as a conjunctive direction-consistency criterion across three correlated replications instead of a single significance test on one designated LLM.

One amendment to the registration was filed on 2026-08-22, while the main collection was running. The review and fix agents read and search the worktree through the small tool surface of \Section{sec:method-instrument}, and two of those tools bounded their result by number of lines and not by number of characters, while the tool returning a diff was not bounded at all. A single line of a minified bundle, a generated source file, or a committed build artifact can be megabytes long, so one tool call could return an arbitrarily large result: in the part of the collection completed when the defect was found, 47 agent sessions across 27 PRs contained at least one call returning 100,000 characters or more, and the largest single result was 11.3 million characters. When an oversized result pushes the next LLM call past the context window, the session terminates as an error; we confirmed four such sessions. When a retry of the same role instead avoids the fatal call, the session completes normally, and the oversized results it did absorb are recorded as ordinary cost: one such DesignReviewer session completed with 1.8 million input tokens, two to three orders of magnitude above a normal session on a comparable diff, with nothing in its record to distinguish it from unaffected data.

The fix caps the searching and file-reading tools at 2,000 characters per line and replaces any longer line with a fixed marker. The threshold was set by measurement: across 283,829 source lines sampled from 55 repositories in the study's own clone cache it blanks 0.022\% of lines, and it is about 150 times shorter than the shortest pathological line confirmed in our data, at 310,647 characters. The diff tool is treated differently and is not capped at all. Its output is written to a session-scoped file inside the worktree and served through the already paginated file-reading tool, so nothing is discarded. We chose a lossless mechanism for the diff tool because PRs that trigger large diff reads skew toward the large-footprint tercile, at 50.0\% against 32.7\% in the full sample, and footprint is both a term and a mediator candidate in the primary model, so truncating at that tool would have introduced an artifact correlated with a predictor of interest.

The amendment changed no agent prompt, no LLM, no sampling decision, and no part of the analysis plan; the prompt hashes filed with the registration are unchanged, and the amendment records the commit at which the corrected instrument was re-frozen. Collection was stopped when the defect was found and resumed only after the amendment had been filed. Of the 465 PRs, 67 had data collected under at least one pre-fix tool, 28 of them showing the oversized-output signature directly; all 67 were archived byte for byte and then collected again from scratch under the corrected instrument, so every estimate in \Section{sec:results} is based on data from the corrected tools and the achieved sample is unchanged. Had the defect gone unnoticed, those PRs would have entered the primary regression with token counts inflated by up to three orders of magnitude, and that inflation is not distributed at random: it occurs when a changed repository happens to contain a large generated or minified file, which depends on the repository and not on the CodeHealth of the files the PR changes, and it concentrates in the largest footprint tercile. We cannot say which way such rows would have moved \(\beta_{\mathit{health}}\). Because they would have sat unmarked in the tail of the primary outcome, we removed the defect at the instrument instead of correcting for it in the analysis.

\subsection{Replication Package}
\label{sec:method-package}

Consistent with community recommendations on open artifacts in empirical software engineering~\citep{runeson_conceptual_2024}, the complete study repository (comprising the pipeline, the agent prompts, the sampler, the analysis scripts, the exploratory content-provenance measurement and its outputs, and the documentation, together with the frozen dataset as pre-registered) is available at \href{https://doi.org/10.5281/zenodo.22127920}{DOI:10.5281/zenodo.22127920}. The pre-registration itself, the prompt hash manifest, the full smell selection surface, the data sample manifest, and the footprint tercile cut points are available as supporting files on OSF \href{https://doi.org/10.17605/OSF.IO/FBJCT}{DOI:10.17605/OSF.IO/FBJCT}.

Each PR's structured session record, comprising per-agent token counts, iteration counts, tool-call counts, termination reasons, emitted comments, the applied patch, and CodeHealth before and after, is included for every run the paper reports, as is the raw transcript of every agent session behind those records: the prompts as sent, every tool invocation in order, the LLM's reasoning and output at each turn, and the guard decision that ended the loop. Transcripts record the size of each tool result, not its content; because the review agents' tools are deterministic functions of the repository state and the logged arguments, and the tool implementation, the repository, and the exact commit are all published, what an agent was shown can be reconstructed by replaying its logged call sequence, with the recorded result sizes as the check. The fix agent is the exception: its edit payloads are elided and its working tree changes as edits are applied, so its intermediate states are bounded by the head commit and the applied patch but cannot be reproduced byte for byte.

\subsection{Use of LLMs and Reporting Guidelines}
\label{sec:method-llm-guidelines}

Beyond the review and fix agents that constitute the study's instrument (\Section{sec:method-instrument}), the software stack used to execute and analyze the experiment, comprising the pipeline orchestration, the sampler, the agentic harness, and the analysis scripts, was developed with the assistance of Claude Code (Anthropic), using Sonnet 5 and Opus 5, for feature planning, code generation, debugging, testing, and code improvement. Anthropic's Opus 5.5 was used to review and critique this paper's draft. The authors take full responsibility for the final text.

The study uses the guidelines for empirical software engineering studies involving large language models of \citet{baltes_guidelines_2025}, which organize such work into seven study types and specify eight guidelines with requirements and recommendations. Under that taxonomy, the four review agents and the fix agent constitute a new software engineering tool built from LLMs. The applicable requirements concern declaration of LLM usage, LLM version and configuration, system and prompt design, session traces, and the inclusion of an open-weight baseline. The paragraph above declares the LLM usage outside the instrument. \Section{sec:method-models} reports each LLM's identifier, version, serving stack, and generation parameters, with the locally served LLM pinned to an exact revision and a hash-checked environment. \Section{sec:method-instrument} describes the agent roles, the control loop, and the tool surface. Every prompt was hashed before registration and is published in full. \Section{sec:method-package} describes the per-PR session records and the raw transcript of every agent session. The locally served open-weight LLM is present alongside the two commercial LLMs, as the guidelines recommend for reproducibility.

Two recommendations are met only in part. \citet{baltes_guidelines_2025} recommend repeating experiments to characterize LLM non-determinism. For cost and environmental impact reasons, the repetition is limited to the 27-PR subsample of \Section{sec:method-replication}. They further recommend validating LLM outputs against human judgment where the quality of the generated artifacts bears on the conclusions; we do not assess whether the review comments are correct or the fixes are good. Both gaps are discussed as threats, under reliability of treatment implementation (\Section{sec:threats-conclusion}) and mono-method bias (\Section{sec:threats-construct}).

\section{Results}
\label{sec:results}

We report the registered analysis on the achieved main collection of 465 PRs, reviewed and fixed once by each of the three LLMs in \Table{tab:models}. Given the chronology in \Section{sec:method-prereg}, every number from \Section{sec:results-attrition} to \Section{sec:results-rq4} is confirmatory in the sense fixed by \Section{sec:method}. The analyses in \Section{sec:results-exploratory}, which close the section, are exploratory and corrected under Benjamini-Hochberg as one family.

\subsection{Sample and Attrition}
\label{sec:results-attrition}

Of the 465 PRs, \texttt{deepseek-v4-flash} completed 453 (97.4\%), \texttt{qwen3-6} completed 461 (99.1\%), and \texttt{sonnet-5} completed 453 (97.4\%), across 243, 243, and 242 distinct repositories respectively. Every excluded PR is attributable to a guard-stopped fix session (\Table{tab:termination}); no PR was excluded for infrastructure failure. Because the three fits are separate per LLM, a PR excluded from one LLM's fit is retained for the other two.

\subsection{\RQ{1}: Baseline CodeHealth and Cost}
\label{sec:results-rq1}

We first apply the decomposition checklist from \Section{sec:experiment-design} to every LLM. The repository-by-cell audit was clean: the largest single-repository share of any tier's PRs was 4.1\%. The single-PR outlier check likewise cleared, with no single PR accounting for more than 8.2\% of any tier's token sum, well under the 20\% alarm threshold. Footprint-adjusted cost (median tokens per changed line) decreased monotonically from the unhealthy to the healthy tier for every LLM, so the tier gradient survived normalizing for PR size. One flag did surface: the mean-to-median ratio for \texttt{sonnet-5} exceeded the 1.5 skew threshold in all three tiers, so we report \texttt{sonnet-5} using medians only and treat its mean token total as uninformative.

\begin{table}[thb]
\caption{The registered primary model, \(\log(\mathit{tokens\_total}) \sim \mathit{health} + \log(\mathit{footprint}) + \mathit{language} + (1|\mathit{repo})\), fitted separately per LLM by restricted maximum likelihood on the achieved main collection. \textbf{Takeaway:} All three coefficients are negative and all three 95\% confidence intervals exclude zero, so \RQ{1}'s conjunctive criterion is satisfied on both its direction clause and its interval clause.}
\label{tab:primarymodel}
\begin{tabular}{lrrrlrr}
\toprule
LLM & \(n\) & Repos & \(\beta_{\mathit{health}}\) & 95\% CI & \(p\) & ICC \\
\midrule
\texttt{deepseek-v4-flash} & 453 & 243 & $-0.062$ & $[-0.090, -0.035]$ & $<.001$ & .221 \\
\texttt{qwen3-6}           & 461 & 243 & $-0.049$ & $[-0.085, -0.014]$ & $.006$  & .032 \\
\texttt{sonnet-5}          & 453 & 242 & $-0.036$ & $[-0.070, -0.001]$ & $.041$  & .131 \\
\bottomrule
\end{tabular}
\end{table}

\begin{figure}[thb]
\centering
\includegraphics[width=\linewidth]{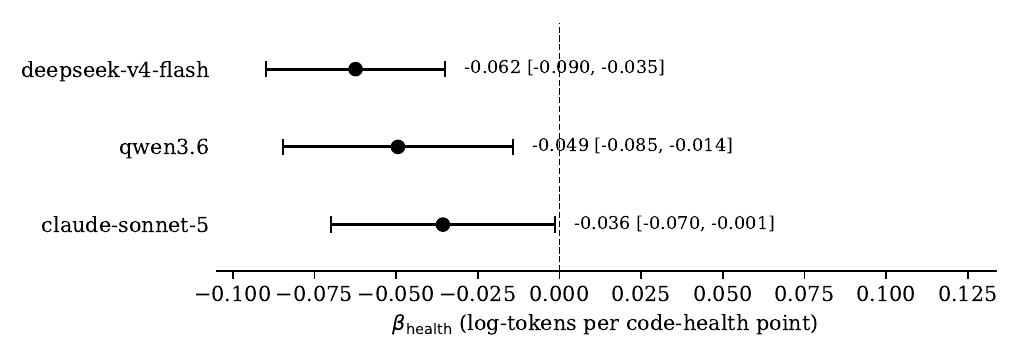}
\caption{Point estimates and 95\% confidence intervals for \(\beta_{\mathit{health}}\), one fit per LLM, from \Table{tab:primarymodel}. \textbf{Takeaway:} All three intervals sit on the negative side of the reference line at zero, and \texttt{sonnet-5}'s interval, whose margin is the narrowest of the three, still clears it, which is why the registered criterion, two of three intervals excluding zero, is met with a margin.}
\label{fig:primarymodelci}
\end{figure}

\Table{tab:primarymodel} and \Figure{fig:primarymodelci} report the primary fit. All three coefficients were negative with 95\% confidence intervals that excluded zero, so H1 is supported. The registered criterion required all three estimates to be negative and at least two of the three intervals to exclude zero; the result exceeded that threshold, since all three excluded it. One additional CH point was associated with an estimated 6.1\% reduction in total tokens for \texttt{deepseek-v4-flash} (95\% CI [3.4, 8.6]\%), 4.8\% for \texttt{qwen3-6} (95\% CI [1.4, 8.1]\%), and 3.5\% for \texttt{sonnet-5} (95\% CI [0.1, 6.8]\%), holding footprint, language, and repository constant. \texttt{sonnet-5}'s interval came closest to zero, which we return to under conclusion validity (\Section{sec:threats-conclusion}).

The median unhealthy-tier PR scores CH 2.8 and the median healthy-tier PR CH 9.7, a difference of 6.9 CH points. We chose to compare the difference between typical PRs for each tier, rather than between the extremes of the 1--10 scale. This gives an aggregated reduction in token cost going from a typical unhealthy-tier PR to an equivalently sized healthy-tier one of 35.0\% for \texttt{deepseek-v4-flash}, 28.9\% for \texttt{qwen3-6}, and 21.8\% for \texttt{sonnet-5}. These figures are algebraic transforms of the fitted point estimate \(\beta_{\mathit{health}}\), not medians or means of the per-PR token totals.

\paragraph{Sensitivity to non-complete sessions.}
\label{sec:results-sensitivity}
The registered sensitivity analysis refits the primary model without the PRs that contain a non-complete agent session (\Section{sec:method-attrition}). For \texttt{deepseek-v4-flash} and \texttt{sonnet-5}, no retained PR contains one, so their sensitivity fits coincide with \Table{tab:primarymodel}. For \texttt{qwen3-6}, excluding the 103 affected PRs attenuated the coefficient from \(\beta = -0.049\) (95\% CI \([-0.085, -0.014]\)) to \(\beta = -0.026\) (95\% CI \([-0.068, 0.016]\), \(p = .23\), \(n = 358\) across 203 repositories). The excluded PRs were not a random subset of the sample: their median baseline CodeHealth was 6.9 against 8.4 for the retained PRs, their median footprint was 134 changed lines against 87, and their median token total was 2.6 times higher. All but one of their 120 guard-stopped sessions ended as a \texttt{spiral}, whereas the pilot evidence in \Section{sec:method-attrition} concerned truncated sessions. Excluding these PRs therefore removes the low-CH, high-cost end of the distribution, i.e., it conditions the sample on a consequence of the outcome, so an attenuated slope is expected. H1 remains supported under the sensitivity fits, with all three coefficients negative and two of three intervals excluding zero. However, \texttt{qwen3-6}'s own interval excludes zero only with these sessions retained, so that LLM's individual result depends on the registered decision to count a spiraling review as cost rather than as noise.

\RQanswer{\RQ{1}}{\emph{Does baseline CodeHealth predict the token cost of agentic review and fix, controlling for PR footprint, language, and repository?} Yes. One CH point is associated with an estimated 3.5\% to 6.1\% fewer total tokens (\Table{tab:primarymodel}).}

\subsection{\RQ{2}: Direct or Footprint-Mediated Effect}
\label{sec:results-rq2}

\RQ{2} asks whether the association in \Section{sec:results-rq1} is direct or whether it runs through the size of the PR that unhealthy code attracts. \Table{tab:rq2paths} reports all three paths of the registered decomposition (\Section{sec:method-analysis}). Two of them are coefficients of the primary fit in \Table{tab:primarymodel}; the CodeHealth-to-footprint path is one further fit per LLM, \(\log(\mathit{footprint}) \sim \mathit{health} + \mathit{language} + (1|\mathit{repo})\), which keeps the registered right-hand side except for \(\log(\mathit{footprint})\) itself, because it is the outcome there.

\begin{table}[t]
\caption{\RQ{2} three-path decomposition, every path estimated in the registered mixed-model family with a repository random intercept, fitted separately per LLM. Path (a) is \(\log(\mathit{footprint}) \sim \mathit{health} + \mathit{language} + (1|\mathit{repo})\); paths (b) and (c) are the \(\log(\mathit{footprint})\) and \(\mathit{health}\) coefficients of the primary fit in \Table{tab:primarymodel}. \textbf{Takeaway:} CodeHealth is not associated with PR footprint for any LLM, footprint is the strongest driver of cost for every LLM, and the CH effect on cost persists with footprint held constant, so the effect is direct with respect to PR footprint; \Section{sec:results-filesize} examines file size.}
\label{tab:rq2paths}
\begin{tabular}{llrlr}
\toprule
Path & LLM & \(\beta\) & 95\% CI & \(p\) \\
\midrule
(a) CodeHealth \(\rightarrow \log(\mathit{footprint})\) & \texttt{deepseek-v4-flash} & $+0.014$ & $[-0.049, +0.078]$ & $.657$ \\
                                                     & \texttt{qwen3-6}           & $+0.014$ & $[-0.050, +0.077]$ & $.672$ \\
                                                     & \texttt{sonnet-5}          & $+0.007$ & $[-0.056, +0.069]$ & $.838$ \\
\midrule
(b) \(\log(\mathit{footprint}) \rightarrow\) cost      & \texttt{deepseek-v4-flash} & $+0.445$ & $[+0.405, +0.484]$ & $<.001$ \\
                                                     & \texttt{qwen3-6}           & $+0.325$ & $[+0.272, +0.378]$ & $<.001$ \\
                                                     & \texttt{sonnet-5}          & $+0.595$ & $[+0.544, +0.647]$ & $<.001$ \\
\midrule
(c) CodeHealth \(\rightarrow\) cost \(\mid\) footprint & \texttt{deepseek-v4-flash} & $-0.062$ & $[-0.090, -0.035]$ & $<.001$ \\
                                                     & \texttt{qwen3-6}           & $-0.049$ & $[-0.085, -0.014]$ & $.006$ \\
                                                     & \texttt{sonnet-5}          & $-0.036$ & $[-0.070, -0.001]$ & $.041$ \\
\bottomrule
\end{tabular}
\end{table}

The CodeHealth-to-footprint path was indistinguishable from zero for every LLM (\(\beta\) between \(+0.007\) and \(+0.014\), every interval spanning zero, \(p \geq .657\)), so unhealthy PRs in this sample were not systematically larger ones. The joint stratification in \Section{sec:method-sampling} was designed to make this path observable, and the data show no such association. The footprint-to-cost path was the strongest association anywhere in the study (\(\beta = +0.325\) to \(+0.595\) per unit of \(\log\) footprint, all \(p < .001\)), confirming that diff size is the dominant driver of token cost. The direct CH-to-cost path, conditional on footprint, is the coefficient already reported in \Table{tab:primarymodel}, negative with an interval excluding zero for all three LLMs. The CH effect is therefore direct with respect to PR footprint, which does not mediate it: unhealthy code was not measurably associated with larger PRs in this sample, and its extra cost persisted after footprint was held constant.

\RQanswer{\RQ{2}}{\emph{Is any such relationship direct, or is it mediated by the size of the PR that
unhealthy code attracts?} Direct with respect to footprint. The CodeHealth effect survives conditioning on footprint (\Section{sec:results-rq2}); its relation to file size is examined in \Section{sec:results-filesize}.}

\subsection{\RQ{3}: Specific Code Smells and Fix Cost}
\label{sec:results-rq3}

\Table{tab:smelltest} tests the three smells selected in \Table{tab:smells} against the main collection, as specified in \Section{sec:method-analysis}.

\begin{table}[htb]
\caption{\RQ{3} confirmatory test on the main collection: partial Spearman correlation with \(\log_{1p}\) fix-agent tokens, controlling for \(\log\) changed lines, Holm-corrected within each LLM's three-hypothesis family. \textbf{Takeaway:} Complex Method and Complex Conditional are confirmed for all three LLMs; Code Duplication is confirmed for none of them, and its sign even reverses for \texttt{sonnet-5}, so only two of the three pre-registered smell hypotheses are supported.}
\label{tab:smelltest}
\begin{tabular}{llrrl}
\toprule
\textbf{LLM} & \textbf{Smell} & \textbf{\(\rho_{\mathrm{adj}}\)} & \textbf{\(p_{\mathrm{Holm}}\)} & \textbf{Confirmed} \\
\midrule
\texttt{deepseek-v4-flash} & Complex Method      & $.153$  & $.002$  & Yes \\
                            & Code Duplication    & $.080$  & $.089$  & No \\
                            & Complex Conditional & $.212$  & $<.001$ & Yes \\ \midrule
\texttt{qwen3-6}            & Complex Method      & $.207$  & $<.001$ & Yes \\
                            & Code Duplication    & $.013$  & $.774$  & No \\
                            & Complex Conditional & $.157$  & $.001$  & Yes \\ \midrule
\texttt{sonnet-5}           & Complex Method      & $.145$  & $.004$  & Yes \\
                            & Code Duplication    & $-.032$ & $.502$  & No \\
                            & Complex Conditional & $.166$  & $.001$  & Yes \\
\bottomrule
\end{tabular}
\end{table}

Complex Method and Complex Conditional showed Holm-significant correlations in all three LLMs, at adjusted correlations between .145 and .212. Code Duplication could not be confirmed for any LLM, and its sign even reversed for \texttt{sonnet-5}, despite its ranking second on the pilot's selection criterion (\Table{tab:smells}). We interpret the pattern in \Section{sec:discussion-smells}.

\RQanswer{\RQ{3}}{\emph{Do specific code smells, rather than the aggregate CodeHealth score, predict the
cost of the fix stage in particular?} The code smells \emph{Complex Method} and \emph{Complex Conditional} predict fix-stage token cost in all three LLMs. \emph{Code Duplication} does not reach significance in any of them (\Table{tab:smelltest}).}

\subsection{\RQ{4}: CodeHealth After Batch Fix}
\label{sec:results-rq4}

\Table{tab:deltahealth} reports \(\Delta\)CodeHealth, the post-fix score minus the baseline score, per LLM. As registered, no hypothesis test is attached to this research question; it is reported as a description.

\begin{table}[htb]
\caption{\(\Delta\)CodeHealth under a single batch fix pass, per LLM. \textbf{Takeaway:} The batch fix leaves CodeHealth essentially unchanged at the median for every LLM, replicating the candidate finding from pilot 3 on the full main collection.}
\label{tab:deltahealth}
\begin{tabular}{lrrr}
\toprule
LLM & \(n\) & Median & Mean \\
\midrule
\texttt{deepseek-v4-flash} & 453 & $0.000$  & $-0.007$ \\
\texttt{qwen3-6}           & 461 & $0.000$  & $-0.026$ \\
\texttt{sonnet-5}          & 453 & $0.000$  & $-0.008$ \\
\bottomrule
\end{tabular}
\end{table}

\begin{figure}[thb]
\centering
\begin{subfigure}[b]{0.32\linewidth}
\includegraphics[width=\linewidth]{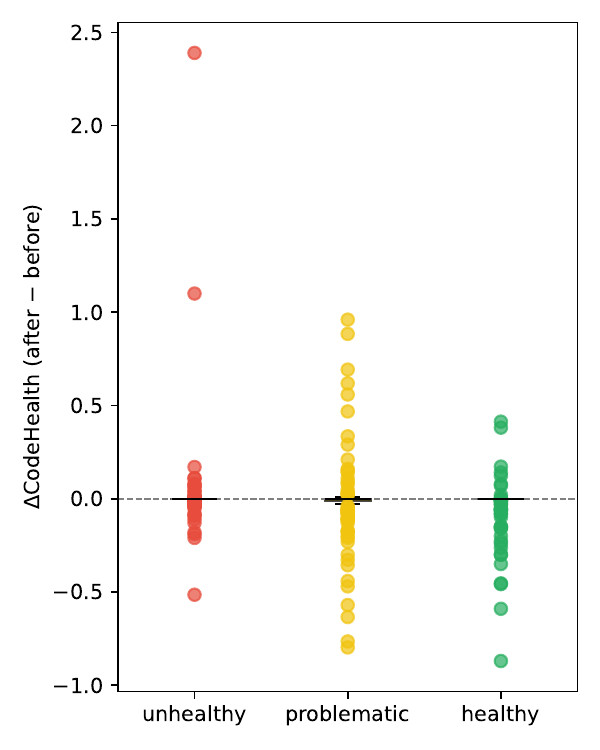}
\caption{\texttt{deepseek-v4-flash}}
\end{subfigure}\hfill
\begin{subfigure}[b]{0.32\linewidth}
\includegraphics[width=\linewidth]{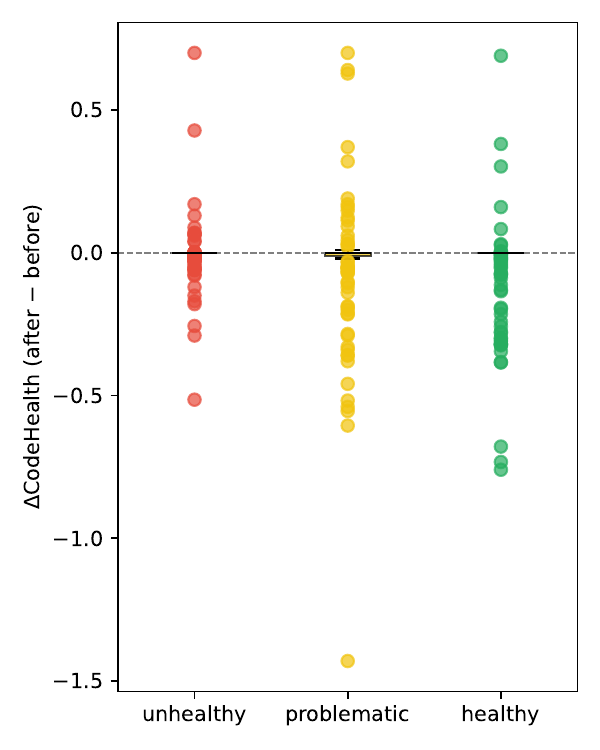}
\caption{\texttt{qwen3-6}}
\end{subfigure}\hfill
\begin{subfigure}[b]{0.32\linewidth}
\includegraphics[width=\linewidth]{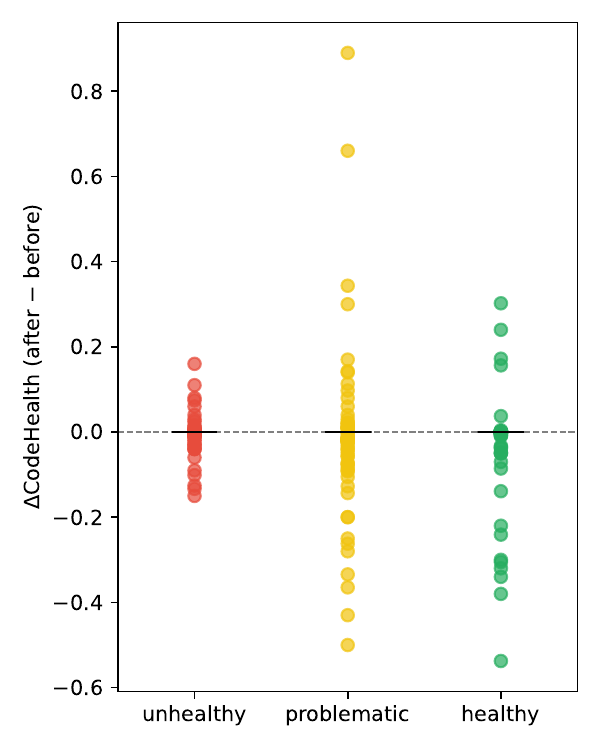}
\caption{\texttt{sonnet-5}}
\end{subfigure}
\caption{Distribution of \(\Delta\)CodeHealth by baseline CodeHealth tier (unhealthy, problematic, healthy, colored red/yellow/green respectively), with the reference line at zero. \textbf{Takeaway:} The mass sits on the zero line in every tier and every LLM, but the spread is two-sided, so the zero median summarizes a distribution in which individual PRs both improve and degrade.}
\label{fig:healthdelta}
\end{figure}

The median \(\Delta\)CodeHealth was exactly zero for every LLM, and the means were small and negative. \Figure{fig:healthdelta} shows the per-LLM distributions behind those medians. A zero median does not mean that the fix pass leaves files untouched: the score was unchanged for 44.4\%, 44.7\%, and 49.0\% of retained PRs, and among those that did move, degradations outnumbered improvements for every LLM, by 158 to 94, 168 to 87, and 136 to 95. Post-fix measurement succeeded for every retained PR, so no \(\Delta\)CodeHealth value is missing. The zero median also held within each baseline tier, for every LLM, as \Figure{fig:healthdelta} shows. Two limits on interpreting it follow from the descriptive form. First, a zero median shows only that no systematic shift is visible. It does not show that the batch fix pass neither repairs nor degrades CodeHealth, since no power analysis was made for \(\Delta\)CodeHealth, and a description cannot exclude a small systematic effect in either direction. Second, the excess of degradations over improvements is untested for the same reason, and its consistency across the three LLMs is weaker support than it appears, since all three LLMs act on the same diffs (\Section{sec:method-analysis}); we do not interpret it as a finding, as it may be a coincidence of this sample.

\RQanswer{\RQ{4}}{\emph{What happens to CodeHealth itself when a batch fix agent applies a full set of
review comments in a single pass?} No systematic shift is visible. Median \(\Delta\)CodeHealth is exactly zero for every LLM, and within every baseline tier. This is a description, and it does not establish that the fix pass leaves CodeHealth unaffected.}

\subsection{Exploratory Analyses}
\label{sec:results-exploratory}

We apply the Benjamini-Hochberg correction once, over a single family consisting of the 42 exploratory hypothesis tests reported in this subsection: the 30 stage and review-stage refits behind \Table{tab:stagedecomp} and \Table{tab:reviewdecomp}, including the matched-subsample refits quoted in the text and the two registered cache-neutral secondary outcomes, the 6 coefficients of \Table{tab:commentdecomp}, the 3 size-adjusted CodeHealth coefficients of \Table{tab:filesize}, and 3 zero-comment selection checks. The family is pooled across the three LLMs, since it counts the exploratory claims the paper makes, not the rows of any one LLM's fit. The registration places every analysis that is not confirmatory under this correction, so the registered secondary outcomes are counted here and are not reported uncorrected. Two sets of tests belong outside it: the confirmatory fits of \Section{sec:results-rq1} and \Section{sec:results-rq2}, which the registration governs, and the nine \RQ{3} smell hypotheses, which are Holm-corrected within each LLM's own registered three-hypothesis family. Every table below reports the unadjusted \(p\) next to the adjusted \(q_{\mathrm{BH}}\), and we call an exploratory result significant only at \(q_{\mathrm{BH}} < .05\). 

\subsubsection{Where the Tokens Are Spent: Pipeline Stages}
\label{sec:results-stages}

We decompose the \RQ{1} effect by pipeline stage to find where in the code review and fix process the CodeHealth association is strongest. The registered outcome, \(\mathit{tokens\_total}\), sums two stages that behave differently: a review stage that runs unconditionally on every PR, and a fix stage that runs only when the reviewer agents produced comments to act on. The analysis in this subsection refits the registered model form, \(\log(\mathit{outcome}) \sim \mathit{health} + \log(\mathit{footprint}) + \mathit{language} + (1|\mathit{repo})\), on each stage separately.

\begin{table}[t]
\caption{The registered model form refitted on each pipeline stage separately, one fit per LLM per outcome, on the same PRs as \Table{tab:primarymodel}. Fix-stage fits drop the PRs with zero review comments: \(n\) reports the rows that remain. \textbf{Takeaway:} The CodeHealth effect survives removing the fix stage from the outcome entirely, and in fact strengthens for every LLM, so the association lies with the review stage and not with the fix stage that follows it.}
\label{tab:stagedecomp}
\footnotesize
\setlength{\tabcolsep}{4pt}
\begin{tabular}{llrrlrrr}
\toprule
LLM & Outcome & \(n\) & \(\beta_{\mathit{health}}\) & 95\% CI & \(p\) & \(q_{\mathrm{BH}}\) & \%/point \\
\midrule
\texttt{deepseek-v4-flash} & Total tokens\textsuperscript{\ddag}  & 453 & $-0.062$ & $[-0.090, -0.035]$ & $<.001$ & --- & $-6.1\%$ \\
                           & Review tokens & 453 & $-0.069$ & $[-0.097, -0.041]$ & $<.001$ & $<.001$ & $-6.7\%$ \\
                           & Fix tokens    & 424 & $-0.034$ & $[-0.067, -0.001]$ & $.041$  & $.064$ & $-3.4\%$ \\ \midrule
\texttt{qwen3-6}           & Total tokens\textsuperscript{\ddag}  & 461 & $-0.049$ & $[-0.085, -0.014]$ & $.006$  & --- & $-4.8\%$ \\
                           & Review tokens\textsuperscript{\dag} & 461 & $-0.059$ & $[-0.097, -0.021]$ & $.002$  & $.007$ & $-5.8\%$ \\
                           & Fix tokens    & 379 & $-0.030$ & $[-0.065, 0.004]$  & $.087$  & $.126$ & $-3.0\%$ \\ \midrule
\texttt{sonnet-5}          & Total tokens\textsuperscript{\ddag}  & 453 & $-0.036$ & $[-0.070, -0.001]$ & $.041$  & --- & $-3.5\%$ \\
                           & Review tokens & 453 & $-0.039$ & $[-0.073, -0.006]$ & $.020$  & $.038$ & $-3.9\%$ \\
                           & Fix tokens\textsuperscript{\dag} & 369 & $-0.054$ & $[-0.094, -0.015]$ & $.007$ & $.018$ & $-5.3\%$ \\
\bottomrule
\end{tabular}
\footnotesize{\textsuperscript{\dag}For these fits the repo random-intercept variance is estimated at zero, leaving the random effect unidentified; they are therefore reported by ordinary least squares with repo-cluster-robust standard errors. The coefficients are unaffected, as the mixed model reduces to least squares when that variance vanishes. \textsuperscript{\ddag}The total-token rows are the confirmatory fits of \Table{tab:primarymodel}, reproduced here for comparison; they belong to the registered analysis and are not FDR-adjusted.}
\end{table}

\Table{tab:stagedecomp} reports the result. Removing the fix stage strengthened the CodeHealth effect. The review-only coefficient was more negative than the total-token coefficient for all three LLMs, moving from \(-0.062\) to \(-0.069\) for \texttt{deepseek-v4-flash}, from \(-0.049\) to \(-0.059\) for \texttt{qwen3-6}, and from \(-0.036\) to \(-0.039\) for \texttt{sonnet-5}. All three review-only intervals exclude zero and all three survive the correction, so the conjunctive criterion registered for \RQ{1} would also be satisfied on a review-only outcome, with slightly more margin than the registered outcome provides. The ratio of coefficient to standard error rises for every LLM as well, from 4.46 to 4.84, from 2.75 to 3.06, and from 2.04 to 2.33, respectively, so the review-only outcome contains the same signal with less noise.

The reason is visible in how the two stages divide the token usage. The review stage accounts for a median of 84.7\%, 86.2\%, and 85.8\% of per-PR tokens for the three LLMs, so the fix stage contributes about one token in seven to the registered outcome. That minority share is also the higher-variance one. The fix stage does not run at all when reviews produced no comments, which is the case for 6.4\% of \texttt{deepseek-v4-flash} PRs, 17.8\% of \texttt{qwen3-6} PRs, and 18.5\% of \texttt{sonnet-5} PRs, and its size is then driven by how many comments were raised rather than by how hard the code was to reason about. The CodeHealth coefficient on fix-stage tokens is the weaker of the two for two of the three LLMs, and it is the one row of \Table{tab:stagedecomp} that the correction demotes, since \texttt{deepseek-v4-flash}'s fix-stage coefficient moves from \(p = .041\) to \(q_{\mathrm{BH}} = .064\) and \texttt{qwen3-6}'s was never significant to begin with. For the third LLM, \texttt{sonnet-5}, the two stages cannot be separated: its review and fix coefficients, \(-0.039\) \([-0.073, -0.006]\) against \(-0.054\) \([-0.094, -0.015]\), have intervals that overlap across most of their ranges.

To rule out that this comparison is an artifact of the larger sample behind the review-stage fit, we refit both stages on the subsample of PRs where the fix stage ran, so that review and fix stages are estimated on an identical sample. The ordering is unchanged for \texttt{deepseek-v4-flash} (\(-0.066\) for review against \(-0.034\) for fix, \(n = 424\)) and \texttt{qwen3-6} (\(-0.062\) against \(-0.030\), \(n = 379\)), and \texttt{sonnet-5} remains indeterminate (\(-0.044\) against \(-0.054\), \(n = 369\), again with overlapping intervals). The stage ordering therefore holds for two LLMs and is undetermined, but not contradicted, for the third.

\subsubsection{Review-Stage Decomposition: Iterations, Input, Output, and Comments}
\label{sec:results-reviewdecomp}

\begin{table}[t]
\caption{Decomposition of the review-stage effect, same model form and same rows as \Table{tab:stagedecomp}. Because \(\log(\mathit{review\ tokens}) = \log(\mathit{iterations}) + \log(\mathit{tokens\ per\ iteration})\), the first two coefficients in each block sum to that LLM's review-token coefficient by construction. Comment counts, and the per-comment ratio, are modeled on the PRs with at least one comment. The last row of each block corresponds to the review-token row of \Table{tab:stagedecomp} minus the sixth row of this block, but it is fitted independently on the ratio, on the smaller sample that having at least one comment requires. \textbf{Takeaway:} Lower CodeHealth is associated with more agent turns and more tokens within each turn, and it shows most consistently on the input side, yet it does not correlate with additional review comments, so the additional spend buys no additional output and the cost per review finding rises accordingly.}
\label{tab:reviewdecomp}
\footnotesize
\setlength{\tabcolsep}{4pt}
\begin{tabular}{llrrlrr}
\toprule
LLM & Review-stage outcome & \(n\) & \(\beta_{\mathit{health}}\) & 95\% CI & \(p\) & \(q_{\mathrm{BH}}\) \\
\midrule
\texttt{deepseek-v4-flash} & Iterations\textsuperscript{\S}            & 453 & $-0.037$ & $[-0.053, -0.021]$ & $<.001$ & $<.001$ \\
                           & Tokens per iteration  & 453 & $-0.032$ & $[-0.046, -0.018]$ & $<.001$ & $<.001$ \\
                           & Input, uncached       & 453 & $-0.045$ & $[-0.063, -0.026]$ & $<.001$ & $<.001$ \\
                           & Output                & 453 & $-0.027$ & $[-0.050, -0.004]$ & $.021$  & $.038$ \\
                           & Uncached input + output\textsuperscript{\S} & 453 & $-0.038$ & $[-0.056, -0.020]$ & $<.001$ & $<.001$ \\
                           & Review comments              & 424 & $+0.010$ & $[-0.012, 0.032]$  & $.373$  & $.447$ \\
                           & Review tokens per comment & 424 & $-0.079$ & $[-0.106, -0.052]$ & $<.001$ & $<.001$ \\ \midrule
\texttt{qwen3-6}           & Iterations\textsuperscript{\dag\S} & 461 & $-0.041$ & $[-0.071, -0.011]$ & $.007$  & $.017$ \\
                           & Tokens per iteration  & 461 & $-0.018$ & $[-0.032, -0.003]$ & $.016$  & $.034$ \\
                           & Input, uncached       & 461 & $-0.028$ & $[-0.049, -0.007]$ & $.008$  & $.019$ \\
                           & Output                & 461 & $-0.031$ & $[-0.059, -0.003]$ & $.029$  & $.048$ \\
                           & Uncached input + output\textsuperscript{\S} & 461 & $-0.030$ & $[-0.050, -0.009]$ & $.004$  & $.013$ \\
                           & Review comments              & 379 & $+0.013$ & $[-0.014, 0.040]$  & $.344$  & $.447$ \\
                           & Review tokens per comment & 379 & $-0.075$ & $[-0.121, -0.028]$ & $.002$  & $.006$ \\ \midrule
\texttt{sonnet-5}          & Iterations\textsuperscript{\S}            & 453 & $-0.026$ & $[-0.047, -0.005]$ & $.016$  & $.034$ \\
                           & Tokens per iteration  & 453 & $-0.014$ & $[-0.029, 0.001]$  & $.070$  & $.104$ \\
                           & Input, uncached       & 453 & $-0.025$ & $[-0.046, -0.003]$ & $.023$  & $.040$ \\
                           & Output                & 453 & $-0.024$ & $[-0.077, 0.029]$  & $.372$  & $.447$ \\
                           & Uncached input + output\textsuperscript{\S} & 453 & $-0.025$ & $[-0.048, -0.002]$ & $.033$  & $.053$ \\
                           & Review comments              & 369 & $+0.008$ & $[-0.016, 0.032]$  & $.528$  & $.599$ \\
                           & Review tokens per comment & 369 & $-0.052$ & $[-0.083, -0.020]$ & $.001$  & $.006$ \\
\bottomrule
\end{tabular}
\footnotesize{\textsuperscript{\dag}Reported by ordinary least squares with repo-cluster-robust standard errors, for the reason given under \Table{tab:stagedecomp}. \textsuperscript{\S}Registered cache-neutral secondary outcome (\Section{sec:method-models}); corrected within the exploratory family like every non-confirmatory test.}
\end{table}

\Table{tab:reviewdecomp} decomposes the review stage along two axes: the number of agent iterations against the tokens consumed per iteration, and the input side against the output side. Lower CodeHealth is associated with more agent iterations for every LLM, with all three intervals excluding zero. It is also associated with more tokens inside each iteration, significantly so for two LLMs and at \(q_{\mathrm{BH}} = .104\) for \texttt{sonnet-5}. The two effects are additive on the logarithmic scale and together reconstruct the review-token coefficient to within 0.001 in log units, the identity being exact in the data but not across separately fitted mixed models with their own variance components. On unhealthy code, the agents take more turns and spend more tokens in each turn. The effect appears on the input side for all three LLMs, whereas on the output side it reaches significance for only two. We therefore interpret the input-side result as the more reliable of the two. We report input excluding cached tokens because it isolates the new input tokens in each iteration, whereas the full input token count would include a multiplication by the number of iterations. The registered cache-neutral sum of the two sides follows the input side: its interval excludes zero for all three LLMs, and it survives the correction for \texttt{deepseek-v4-flash} and \texttt{qwen3-6}. For \texttt{sonnet-5} it does not, at \(q_{\mathrm{BH}} = .053\).

One row does not follow this pattern: CodeHealth does not predict how many comments the reviewer agents write. The coefficient is small, positive, and far from significance for any of the LLMs, at \(q_{\mathrm{BH}} \geq .44\). The agents report a comparable number of findings on unhealthy and healthy code, but reaching those findings takes more work on unhealthy code.

Since the work rises but the findings do not, the two can be composed into a single quantity: the review tokens spent per review comment produced. The last row of each block in \Table{tab:reviewdecomp} fits the registered model form on \(\log(\mathit{review\ tokens} / \mathit{comments})\), and it is a re-expression, not a fourth finding, since \(\log(\mathit{review\ tokens}/\mathit{comments}) = \log(\mathit{review\ tokens}) - \log(\mathit{comments})\) and the coefficient is correspondingly the difference of two coefficients already in the table. We report it separately because it is the quantity a practitioner pays for and the most consistent form the effect takes anywhere in this study: one CH point is associated with \(7.6\%\), \(7.2\%\), and \(5.0\%\) lower review cost per finding for the three LLMs, all three intervals exclude zero, and all three reach \(q_{\mathrm{BH}} \leq .006\), including \texttt{sonnet-5}, whose review-token fit is the weakest in \Table{tab:stagedecomp} at \(q_{\mathrm{BH}} = .038\). Across the 6.9-point median-to-median tier contrast used in \Section{sec:results-rq1}, a healthy-tier PR costs an estimated \(42.0\%\), \(40.2\%\), and \(30.0\%\) less inference per review finding than an equivalently sized unhealthy one. The ratio is undefined on PRs where the reviewers raised no comments, which is why it is fitted on the same reduced samples as the comment counts. That exclusion does not bias the slope: whether a PR receives zero comments is unrelated to its baseline CH in all three LLMs (point-biserial \(|r| \leq .022\), \(q_{\mathrm{BH}} \geq .66\)), so the dropped rows are not drawn preferentially from either end of the predictor.

\subsubsection{Code Smell Count and Review Comments}
\label{sec:results-findings}

To test an objection to the comment-count null, we separate the CodeHealth score from the smell count behind it. The objection is that CH is low precisely because CodeScene identified code smells in the files under review, so if the agents react to the same properties, the number of findings should increase in unhealthy code. That they do not could have two explanations: either the agents attend to something largely unrelated to what the metric measures, or the metric is the wrong operationalization of what they attend to.

The distinction that separates these explanations is between the score and the count behind it. CodeHealth is a bounded and saturating index on a 1 to 10 scale, whereas the number of smell instances CodeScene reports is an unbounded count. The two are related but far from interchangeable in our sample, correlating at Spearman \(\rho = -.72\). Within the unhealthy tier alone, where the score is close to its floor for every PR, the smell count still ranges from 4 to 70 with a median of 9. A file can therefore accumulate a great many additional smells without the score moving, which means that a lack of correlation between review comments and the CH score does not by itself rule out a correlation with code smells.

\begin{table}[thb]
\caption{Exploratory refit of the comment-count outcome with both the CodeHealth score and the smell count as predictors on the same rows, controlling for \(\log\) changed lines, \(\log\) scored-file lines, \(\log\) changed files, language, and a repo random intercept. Comment counts are modeled on the PRs with at least one comment, as in \Table{tab:reviewdecomp}. \textbf{Takeaway:} The number of findings the agents report tracks how many smells CodeScene counted, while remaining unrelated to the CH score those smells produce, so the null in \Table{tab:reviewdecomp} reflects the predictor and is not evidence that the agents attend to unrelated properties of the code.}
\label{tab:commentdecomp}
\begin{tabular}{llrrlrr}
\toprule
LLM & Outcome & \(n\) & \(\beta\) & 95\% CI & \(p\) & \(q_{\mathrm{BH}}\) \\
\midrule
\multicolumn{7}{l}{\emph{Smell count}, \(\beta_{\log(1+\mathit{smells})}\)} \\
\texttt{deepseek-v4-flash} & Review comments        & 424 & $+0.210$ & $[+0.104, +0.315]$ & $<.001$ & $<.001$ \\
\texttt{qwen3-6}           & Review comments        & 379 & $+0.249$ & $[+0.120, +0.377]$ & $<.001$ & $.001$ \\
\texttt{sonnet-5}          & Review comments        & 369 & $+0.142$ & $[+0.025, +0.258]$ & $.017$  & $.034$ \\ \midrule
\multicolumn{7}{l}{\emph{CodeHealth score}, \(\beta_{\mathit{health}}\), same fits} \\
\texttt{deepseek-v4-flash} & Review comments        & 424 & $+0.019$ & $[-0.012, +0.049]$ & $.224$  & $.304$ \\
\texttt{qwen3-6}           & Review comments        & 379 & $+0.027$ & $[-0.011, +0.064]$ & $.159$  & $.223$ \\
\texttt{sonnet-5}          & Review comments        & 369 & $+0.008$ & $[-0.026, +0.043]$ & $.629$  & $.665$ \\
\bottomrule
\end{tabular}
\end{table}

\Table{tab:commentdecomp} fits both predictors on the same rows, so that neither can stand in for the other. The smell count predicts the number of review comments for all three LLMs, with every interval excluding zero and every coefficient surviving the correction, while the CodeHealth correlation remains null in the same fits at \(q_{\mathrm{BH}} \geq .22\). In interpretable terms, doubling the number of smells in the files under review is associated with \(15.7\%\), \(18.8\%\), and \(10.3\%\) more review findings for the three LLMs, whereas a full point of CH is associated with at most \(2.7\%\) and never reaches significance. One specification choice favors the smell result: the coefficient is about half as large when the size of the scored files is left out of the model, because larger files contain more smells while drawing fewer comments per file, so file size acts as a suppressor here. We keep file size in the model because the smell count is a count over those files, and without it the fit cannot separate more smells from more code.

\subsubsection{Size of the Scored Files}
\label{sec:results-filesize}

The registered model adjusts for the number of changed lines but not for the size of the files those lines belong to, and the CodeHealth tiers differ greatly in that size: the median scored file has 2,146 lines in the unhealthy tier against 128 in the healthy tier. We therefore refit the registered model with one added covariate, the logarithm of the total lines of the files CH scored for the PR, per LLM and on the same rows as \Table{tab:primarymodel}. We use total lines rather than source lines of code because the agents' file-reading tool pages through a file by line, so comments and blank lines cost tokens just as code does. If the CH coefficient kept its sign with an interval excluding zero, the association would not be reducible to the size of the files the agents read. If it attenuated, we would interpret the effect as mediated by file size, because the size-related smells, such as Large Method, Low Cohesion, and Bumpy Road, are part of what CodeHealth measures (\Section{sec:threats-construct}).

\begin{table}[t]
\caption{The registered model refitted with \(\log\) total lines of the scored files added, one fit per LLM on the same rows as \Table{tab:primarymodel}. The size coefficient is the change in log tokens per unit of log file lines; it is reported as an estimate and is not part of the corrected family. \textbf{Takeaway:} Adjusting for file size absorbs the CodeHealth association for all three LLMs, which in this sample cannot be separated from the size of the files it is computed over.}
\label{tab:filesize}
\footnotesize
\setlength{\tabcolsep}{3pt}
\begin{tabular}{llrrlrrl}
\toprule
LLM & Fit & \(n\) & \(\beta_{\mathit{health}}\) & 95\% CI & \(p\) & \(q_{\mathrm{BH}}\) & \(\beta_{\log \mathit{size}}\) [95\% CI] \\
\midrule
\texttt{deepseek-v4-flash} & Registered\textsuperscript{\ddag} & 453 & $-0.062$ & $[-0.090, -0.035]$ & $<.001$ & --- & --- \\
                           & + file size & 453 & $-0.009$ & $[-0.044, 0.026]$ & $.618$ & $.665$ & $+0.179$ $[0.104, 0.254]$ \\ \midrule
\texttt{qwen3-6}           & Registered\textsuperscript{\ddag} & 461 & $-0.049$ & $[-0.085, -0.014]$ & $.006$ & --- & --- \\
                           & + file size & 461 & $-0.023$ & $[-0.071, 0.025]$ & $.352$ & $.447$ & $+0.082$ $[-0.021, 0.185]$ \\ \midrule
\texttt{sonnet-5}          & Registered\textsuperscript{\ddag} & 453 & $-0.036$ & $[-0.070, -0.001]$ & $.041$ & --- & --- \\
                           & + file size & 453 & $+0.018$ & $[-0.027, 0.064]$ & $.430$ & $.501$ & $+0.171$ $[0.074, 0.268]$ \\
\bottomrule
\end{tabular}
\footnotesize{\textsuperscript{\ddag}The confirmatory fits of \Table{tab:primarymodel}, reproduced for comparison; they are not FDR-adjusted.}
\end{table}

\Table{tab:filesize} reports the result: the association attenuated, the second of the two outcomes described above. Adjusting for file size absorbs the CodeHealth association for every LLM. The coefficient moves from \(-0.062\) to \(-0.009\) for \texttt{deepseek-v4-flash}, from \(-0.049\) to \(-0.023\) for \texttt{qwen3-6}, and from \(-0.036\) to \(+0.018\) for \texttt{sonnet-5}, and none of the three adjusted intervals excludes zero (\(q_{\mathrm{BH}} \geq .44\)). File size itself has a positive coefficient for all three LLMs, with intervals excluding zero for two of them: doubling the scored lines is associated with approximately 13\% more tokens for \texttt{deepseek-v4-flash} and \texttt{sonnet-5}, and 6\% for \texttt{qwen3-6}. The two predictors overlap strongly, with Spearman \(\rho\) between CH and log file size of \(-.71\) to \(-.73\) across the three LLMs' samples, about as strongly as CH correlates with the smell count (\(\rho = -.72\), \Section{sec:results-findings}), so that little variation in CH remains at constant file size, and the adjusted standard errors are 27\% to 37\% wider than the registered ones. The refit shows that this sample cannot separate an association with CH from an association with the size of the files CodeHealth is computed over. It does not by itself show the registered association to be spurious; we analyze this further in \Section{sec:discussion-filesize}.

\subsubsection{Instrument Stability: Run-to-Run Variance}
\label{sec:results-lambda}

\begin{table}[thb]
\caption{Run-to-run variance decomposition on the 27-PR replication subsample (three runs per PR per LLM). \(\lambda\) is the run-to-run variance divided by the between-PR variance; ICC(\(k=1\)) is the implied reliability of a single run. \textbf{Takeaway:} \texttt{qwen3-6}'s run-to-run noise is about an order of magnitude larger, relative to its stable PR signal, than either of the other two LLMs, while \texttt{deepseek-v4-flash} and \texttt{sonnet-5} are both highly reliable at a single run.}
\label{tab:lambda}
\begin{tabular}{lrrrrr}
\toprule
LLM & PRs complete & Var(between) & Var(within) & \(\lambda\) & ICC(\(k=1\)) \\
\midrule
\texttt{deepseek-v4-flash} & 25/27 & 0.939 & 0.052 & 0.055 & .948 \\
\texttt{qwen3-6}           & 27/27 & 0.713 & 0.313 & 0.440 & .695 \\
\texttt{sonnet-5}          & 27/27 & 2.390 & 0.095 & 0.040 & .962 \\
\bottomrule
\end{tabular}
\end{table}

The replication subsample described in \Section{sec:method-replication} partitions each LLM's residual variance into a between-PR component and a run-to-run component, reported in \Table{tab:lambda}. \Section{sec:threats-conclusion} discusses what it means for the interpretation of the confirmatory estimates in \Table{tab:primarymodel}. \texttt{deepseek-v4-flash} and \texttt{sonnet-5} both reach a single-run reliability above .94, so for these two instruments the great majority of variance in a PR's measured cost is a stable property of the PR itself and not of the particular run. \texttt{qwen3-6} is less reliable at a single run, at ICC\((k=1) = .695\), with run-to-run variance nearly half as large as the stable between-PR variance (\(\lambda = 0.440\)). Two PRs are excluded from the \texttt{deepseek-v4-flash} decomposition because their main-collection run did not complete cleanly; the remaining 25 of 27 still supply 50 degrees of freedom on the run-level variance component.

\section{Discussion}
\label{sec:discussion}

A quality metric calibrated against human maintainability predicts what LLM agents spend on reviewing code. The spend arises when the agents read the code and not when they change it, and in our sample it is associated with the size of the files in which unhealthy code occurs. The token cost per finding also increases with decreasing CodeHealth, since the agents write the same number of comments for healthy and unhealthy code while spending more tokens on the unhealthy code. However, in absolute terms the token cost effect is modest. Applied to the 6.9-point median-to-median CH difference from \Section{sec:results-rq1}, the contrast corresponds to about 91,000 tokens per PR for \texttt{sonnet-5}, 134,000 for \texttt{qwen3-6}, and 360,000 for \texttt{deepseek-v4-flash}, for a single review-and-fix pass. The saving recurs with volume: every PR that an organization reviews incurs the same percentage again.

A more capable LLM might be insensitive to code quality, which would make the association a property of current LLMs and not of the code. Our data does not indicate this. The most capable LLM we ran, \texttt{sonnet-5}, still showed a negative coefficient with an interval excluding zero (\Table{tab:primarymodel}), and the ordering of the per-LLM coefficients reversed between pilot 3 and the main collection, with \texttt{sonnet-5} steepest at \(-0.071\) in the pilot (\Table{tab:pilot}) and shallowest at \(-0.036\) in the main collection, so the ordering cannot be taken as a hint about capability. The registered analysis excludes a health-by-LLM interaction term because three LLMs cannot separate capability from the other ways in which they differ (\Section{sec:method-analysis}), as a test of capability would have required a design testing orders of magnitude more LLMs and selecting LLMs with similar architectures differing in size and capability. All three LLMs are current-generation, but none is its provider's flagship tier. Whether the largest and most capable LLMs, at the time of writing, LLMs such as Opus 5.5 or GPT-6 Astra, show the same association remains open.

\subsection{Cost Associated with Code Reading}
\label{sec:discussion-cost}

The exploratory decompositions in \Section{sec:results-stages} and \Section{sec:results-reviewdecomp} locate the association in the review stage. Removing the fix stage steepens the association for every LLM. The association is stronger in the review stage than in the fix stage for two of the three, and the two stages cannot be separated for \texttt{sonnet-5} (\Table{tab:stagedecomp}). We interpret the fix stage as diluting the effect: it runs only when the reviewers produced comments, and its size then follows the number of comments raised rather than the difficulty of the code, so folding it into the registered outcome adds variance that CodeHealth does not explain. The registered \(\mathit{tokens\_total}\) is therefore, if anything, a conservative statement of the association for the part of the pipeline that reads the code, since every LLM's review-only coefficient is steeper than its total-token coefficient. Because the association lies in the review stage, we believe that every additional agent pass that reads the same code, whether a triage step, an explanation, or agent work outside review, could incur a similar token cost association.

\subsection{File Size as a Pathway}
\label{sec:discussion-filesize}

The file-size refit in \Section{sec:results-filesize} changes how the registered result should be interpreted, though not whether it holds. Adjusting for the total lines of the scored files absorbs the CodeHealth coefficient for all three LLMs, and in this sample CH and file size are close to interchangeable descriptions of the same PRs (Spearman \(\rho\) between \(-.71\) and \(-.73\)). We interpret this as the association between CH and cost being mediated by file size, not as a spurious association. For one thing, CH directly uses source lines of code (SLOC) as one of its metrics. We also believe that a large source file is mostly a consequence of many other design choices that lower CodeHealth. Breaking a long file into cohesive functions and classes removes, for example, the Large Method, Low Cohesion, and Bumpy Road smells that the metric penalizes, and once code is split into separate units there is rarely a reason to keep them in one file. Adjusting for file size therefore removes at least one direct factor that the metric measures and several indirect ones, which is why we read the size-adjusted fit as a decomposition of the association and not as a correction of it.

The mediation account is consistent with the review-stage decomposition. An agent reading a long file through a tool that returns at most 500 lines per call (\Section{sec:method-instrument}) needs more calls, and therefore more turns, to locate the context a PR depends on, and each turn adds more retrieved code to its context, which is the pattern of more iterations and more uncached input in \Table{tab:reviewdecomp}. The saving that the estimate in \Section{sec:results-rq1} describes is the one associated with moving from a typical unhealthy-tier PR to a typical healthy-tier one, and such PRs differ in the size of the files they touch as much as in their score, so the estimate does not say what raising CodeHealth while leaving file size unchanged would save. The refactoring that would deliver the saving is the kind that also shrinks files, splitting long methods and low-cohesion modules into smaller units, rather than a change that raises the score without altering the structure an agent has to read through. This interpretation has a limit that we cannot remove with this sample: files that are long for reasons unrelated to design, such as generated code, data tables, or test files with many small cases, would add cost through size without the defects the metric penalizes, and our data cannot separate them from long, low-cohesion files. We return to this question in \Section{sec:discussion-future}.

The file-size result also qualifies how our results converge with \citet{wirdemann_code_2026}, who reported that lower-CodeHealth files require more input tokens in single-shot unit-test generation, for Java and C++ even after controlling for SLOC. Unlike theirs, our association does not survive an adjustment for the size of the files read, although the adjustments differ slightly: they controlled for SLOC, and we controlled for the total lines of all files scored for a PR. That study and ours differ in task, corpus, unit of analysis, and outcome, yet both locate the cost on the input side of the LLM, which we believe indicates that the association originates in the code and not in either pipeline.

\subsection{Cognitive Load and Agentic Inference Cost}
\label{sec:discussion-inferential}

Inside the review stage, lower CodeHealth is associated with more turns taken, more context retrieved into those turns, and more tokens generated within them, while the number of findings the agents report does not change, so the cost per finding rises as CH falls (\Section{sec:results-reviewdecomp}). The intuitive interpretation of this pattern is that unhealthy code is harder for the agents to understand. We do not adopt this interpretation, because it attributes a human comprehension process to the agent. An LLM agent does not comprehend a diff in the sense a human reviewer does, with a mental model assembled over time, prior familiarity with the module, and recognition-primed pattern matching~\citep{gullstrand_heander_code_2026}. It computes over tokens and infers, and both the computing and the inferring are metered. We therefore define what we report by what the instrument measures, namely the turns, retrieved context, and generated tokens listed above, and make no claim about an internal state of understanding.

Cognitive load theory offers a vocabulary and a documented set of symptoms for the human side of this comparison. \citet{sweller_cognitive_1988} showed that conventional problem solving consumes limited cognitive processing capacity, leaving little for acquiring the domain-specific knowledge that distinguishes experts from novices, and \citet{sweller_cognitive_1998} separated that load into load intrinsic to the material, load imposed by how it is presented, and load spent building durable understanding of it. In code review specifically, \citet{goncalves_explicit_2022} treated cognitive load as the mechanism behind review cost and tested whether explicit strategies could lower it, while \citet{siegmund_understanding_2014} located the working-memory and attentional demands of program comprehension in measurable brain activity, which is evidence that the load is a measurable property of the comprehension process. The behavioral signature of that load is well documented. Reviewers re-read a change, navigate to definitions, tests, and history that the diff does not show them, ask clarifying questions when a change fails to make sense, and take longer overall, with \citet{pascarella_information_2018} cataloging the information needs that drive the navigation and \citet{ebert_exploratory_2021} the confusion that drives the questions. 

Our analysis finds a corresponding signature in an agent: additional iterations correspond to re-reading and revisiting, additional context retrieved into those turns to a reviewer's navigation, and additional tokens spent within a turn to the effort of working through a passage. We nonetheless decline to reuse the label \emph{cognitive load} for agents. \citet{shang_united_2025} make that move, mapping cognitive load theory's bounded working memory onto an LLM's context budget by direct analogy, and the analogy commits to a claim our instrument cannot support: that the difficulty arises from an exhausted resource standing between an unprocessed representation and a stable schema. Nothing we measure speaks to whether such a resource exists or is exhausted. For most PRs, our pipeline uses only a fraction of the LLM's context limit. We therefore report the expenditure itself and leave open whether it placed load on any limited resource of the agent, since that is not something we can observe.

Instead of a comprehension account, we propose that interleaved conditional logic, long methods, and low cohesion each force an agent to pull in more context and to trace more paths before a finding can be stated with confidence, without changing how many findings there are to state.

\subsection{Smell Count Versus CodeHealth Score}
\label{sec:discussion-smells}

\Section{sec:results-findings} answers the objection that the comment-count null invites: the number of review findings follows the count of smells in the files under review for all three LLMs, while the CodeHealth score in the same fits does not predict it (\Table{tab:commentdecomp}). The most plausible interpretation, we believe, is that the agents respond to the number of distinct problem sites in front of them, not to how far those problems have moved a summary score. A count and a saturating index diverge when the count grows large, which is where our unhealthy tier sits, and a review agent producing one comment per site it judges worth reporting would produce a comment distribution that follows the count. This interpretation aligns with what we observe in \Section{sec:discussion-inferential}. Cost per finding rises as the score falls, and finding count rises as the smell count grows, so the two results describe different predictors acting on different outcomes.

The review agents are instructed to flag only issues introduced or exposed by the PR under review and explicitly not to flag the unchanged code (\Section{sec:method-instrument}). Unchanged code makes up most of the input files (see provenance shares in \Section{sec:threats-construct}), especially in the unhealthy tier. The agents are consequently instructed not to comment on most of the code responsible for a low score, yet their finding counts still track the count of smells in those files. We interpret this as an indication that the changed code in the PR inherits the smell density of its file, since it needs to interface with the surrounding file and follow similar conventions, although our design cannot confirm that explanation.

The registered smell results (\Section{sec:results-rq3}) are consistent with the structural account of \Section{sec:discussion-inferential} for two of the three smells. Complex Method and Complex Conditional multiply the execution paths a fix agent must trace before it can safely edit, whereas the smell count in aggregate measures how many places invite a comment at all. The Code Duplication prediction failed, and we report it as a failed prediction without reinterpreting its mechanism after the fact.

\subsection{Roads to Code Improvement}
\label{sec:discussion-roads-to-health}

For practitioners who hold refactoring and maintenance budgets, the result is an association, not a forecast. Code that CodeHealth rates as unhealthy also costs more for an agent to review, on every pass an agent makes over it, and in our sample that cost travels with the large, low-cohesion files that such code tends to come in. We believe the case that human maintainability already makes for refactoring can be extended to agent-reviewed code, provided that the refactoring changes the structure of the code and not only its score. Two practices can raise and maintain CH, and a third was considered and rejected for this study.

Metric-guided refactoring improves existing code in passes targeted by the metric instead of by intuition, whether the work is performed manually by a developer, delegated to an agent, or done collaboratively with a developer deciding what to change and an agent proposing and drafting the edits. We favor the collaborative form, for two reasons. The metric shows where improvements can be made but the team that maintains the code should make the final decision on which functions to refactor. Furthermore, \citet{borg_code_2026} report that LLMs preserve program semantics less reliably when working with low-CodeHealth code, with break rates rising as health falls, so an agent operating unsupervised is least dependable in the files that need refactoring the most. We believe metric-guided refactoring should be paired with human review of the resulting change and with test coverage sufficient to detect regressions, rather than run as an unattended batch job.

Another strategy is a quality gate on incoming changes, similar to test coverage and regression gates that many projects already run in continuous integration: a change that lowers the aggregate CodeHealth of the files it touches, or that leaves those files below an agreed target, cannot merge until its score is brought up. This is relatively easy to implement, since the metric is deterministic static analysis and is cheap and fast to compute automatically for every changed file. Because the returns on maintainability appear to increase toward the healthy end of the scale~\citep{borg_increasing_2024}, a gate set at a target, and not merely at no-regression, has a defensible economic argument behind it. Combining this with metric-guided refactoring would give a strategy where the gate holds a line while refactoring moves it. Using only the gate has the advantage of amortizing the cost over the average life cycle of a file in the project, instead of refactoring files that otherwise might not have been touched or reviewed for a long time. The drawback is that some parts of the codebase might be left in an unhealthy state for a long time, eventually putting limits on how good the rest of the codebase can become since it still has to interface with the low-quality remnants.

A third option, giving the fix agent a CodeHealth target, was a candidate for this study, and we decided against it (\Section{sec:experiment-design}). The fix agent in our implementation receives only review comments and no CH target or retry loop (\Section{sec:method-instrument}). We considered providing it with the CodeHealth tool and instructing it to raise the score of the files that it touches. That instruction would guard against code review fixes degrading the health of the project, which the batch fix in our pipeline does to some PRs while improving others (\Section{sec:results-rq4}), and would make each new PR raise the CH score as far as it could. However, we concluded that doing so would risk degrading the change. A small, well-defined bug fix or feature could easily come bundled with dozens of unrelated refactorings, and the reviewer of that pull request would then be asked to evaluate two changes at once, of which only one was requested.

Keeping the two concerns apart also matters after the change has merged. If either the requested change or the refactoring that came with it introduces a defect, the team needs to revert one without losing the other, and a single commit that contains both offers no clean way to do so. A separate commit whose stated purpose is to raise CodeHealth can be reviewed and reverted on its own, and it keeps the commit history an accurate record of why each change was made. We believe that quality-improving refactoring is better delivered as its own documented change than as a side effect of another.

Whichever road is taken, we do not interpret the result as an argument for optimizing code for agents. CodeHealth is calibrated against human maintainability~\citep{tornhill_code_2022, borg_increasing_2024}, and the finding here is that a metric built for human reviewers also predicts agent inference cost. That alignment makes refactoring a shared investment for human and agent readers, and we believe it should be preserved deliberately. If future quality metrics were tuned to minimize agent token cost directly, they could drift away from what makes code tractable for the humans who still own it and are responsible for it. In such a scenario, the interpersonal and knowledge-transfer benefits of review that we have argued for elsewhere~\citep{heander_support_2025} would be at risk.

\subsection{Future Work}
\label{sec:discussion-future}

The first open question is a qualitative review of multi-agent review comments, in comparison to the quality of the code under review. This study counted review comments but did not grade them. \Section{sec:results-reviewdecomp} shows that agents on unhealthy code take more turns without raising more comments, but not whether the comments they \emph{do} raise are equally correct and equally severe. Grading each agentic review comment against its diff, for correctness and for whether it identifies a problem worth fixing, would turn tokens per comment into tokens per useful finding. It would also test whether unhealthy code draws the same number of comments but different ones, for instance, higher-severity comments pointing to architectural issues.

Future work could also reverse the direction of the smell analysis. CodeHealth and its smells were calibrated against human maintenance effort, and nothing guarantees that every smell costs an agent what it costs a human; some constructs penalized for human readers might be neutral or even better for an agent that retrieves context through tools instead of holding it in working memory. A design that isolates individual smells, for example by varying one smell at a time in otherwise matched code and measuring inference cost per finding, would separate the smells that drive the agent cost from those that only co-occur with them, and would say whether a quality metric calibrated for agents would weight the smells differently from one calibrated for humans.

Another direction for a future study is to explore the connection between CodeHealth, file size, and inference cost in more detail. By building a dataset stratified on file size, it could be possible to isolate the effect of larger files from the effect of lower code quality within those files. Exploring different implementation and parameter choices for the tools given to the LLMs could also have large effects on how they explore their context and on how severely larger files increase token cost.

Finally, \RQ{3} was registered on the fix stage, where the agent edits the code. We did not test the smells against review-stage or total cost, which is a natural follow-up, although it would require more data to reach significance over so many input variables.

\section{Threats to Validity}
\label{sec:threats}

We organize the threats to validity by Cook and Campbell's four categories as presented by \citet{wohlin_experimentation_2024}: internal, external, construct, and conclusion validity. Following \citet{lago_threats_2024}, most countermeasures below are sampling, instrumentation, or registration decisions already described in \Section{sec:method}; we describe below what they address and what they leave standing. Cook and Campbell's labels refer to a treatment; in this observational study the role of the treatment is played by the CodeHealth tier, and labels such as \emph{selection by treatment} are to be understood in that sense.

\subsection{Internal Validity}
\label{sec:threats-internal}

\paragraph{Confounding by repository.} PRs from the same repository share conventions, architecture, language, and authorship, so whatever is peculiar to a dominant repository can masquerade as an effect of CodeHealth; the AIDev pool is 39.3\% one repository, and this confound invalidated the second of our pilot samples (\Section{sec:experiment-design}). The repository caps limit any repository to two PRs per cell and nine overall, and the random intercept discounts within-repository replication. The within-repository effect is, however, estimable only from the 55 repositories (22\%) that span at least two tiers.

\paragraph{Overcontrol by PR footprint.} Adjusting for the size of the PR is necessary, because token cost scales mechanically with diff size, and dangerous, because if unhealthy code forces larger PRs then adjusting footprint away would remove part of the effect of interest. We therefore treat footprint as a mediator candidate and report the full three-path decomposition for \RQ{2}. The residual threat is inferential, not statistical: tasks are not matched across CodeHealth levels, so the CH-to-footprint path does not support a causal interpretation.

\paragraph{Instrumentation drift and instrument changes.} The proprietary CodeHealth tool's scoring algorithm could change during a multi-week collection; we pinned the tool and verified that all 465 manifest entries record a single version, 1.4.0. The harness changed three times. Before the freeze, the locally served LLM moved to a new serving stack and two defects that could record an infrastructure failure as a success were fixed; all 45 pilot PRs were re-run, so only the pooling and interaction analyses that motivated parts of the design (\Section{sec:method-analysis}), neither a registered outcome, depend on the superseded stack. After the freeze, unbounded tool output was found to let one oversized result inflate a session's token count without marking its record (\Section{sec:method-prereg}). Collection was stopped, the fix was filed as an amendment before collection resumed, and all 67 affected PRs were collected again from scratch, so no reported estimate comes from a defective instrument. The residual threat is an undetected defect of the same class, which the termination reason published for every agent session (\Table{tab:termination}) would make visible.

\paragraph{Regression to the mean.} PRs entering the pipeline with low baseline CodeHealth have more room to improve, which would inflate \(\Delta\)CodeHealth in the unhealthy tier under any treatment. We do not correct for this, because \RQ{4} is registered as a description, not a hypothesis test, and we report \(\Delta\)CodeHealth by tier so that the asymmetry in headroom is visible.

\subsection{External Validity}
\label{sec:threats-external}

\paragraph{Selection by treatment.} The sample is drawn from merged, agent-authored PRs in repositories with at least 100 stars, in five languages, dominated by one originating agent (OpenAI Codex, 49\%). The findings speak to agentic review of agent-authored PRs in popular open-source projects, the setting where such review is already routine~\citep{he_writes_2026}, and we do not claim they transfer to human-authored PRs, to private or legacy codebases, or to other languages.

\paragraph{Scarcity in the stratum of most interest.} The two cells that fell short of target are both in the unhealthy tier (\Table{tab:cellfill}). The cells where a CodeHealth effect on cost would be most visible are also the thinnest: any claim about medium and large unhealthy PRs rests on 73 PRs instead of 112.

\paragraph{Setting by treatment.} The instrument is our own four-agent pipeline, not a deployed product, and production systems differ in ways that could change the cost profile, for example by pairing a rule-based detector with an LLM filter stage~\citep{sun_bitsai_2025}. Pipeline design can also moderate how code metrics relate to agent outcomes: \citet{ramirez_exploring_2025} found the Maintainability Index correlating with issue resolution at \(r = .103\) and \(r = .436\) for the same LLM, Claude 3.5 Sonnet, under two different scaffolds. We ground the agent partition in ISO/IEC 25010~\citep{iso_25010_2023} and publish the prompts in full, but the results characterize this pipeline and generalize to others only by argument.

\paragraph{History by treatment.} LLM capability, pricing, and serving efficiency move quickly, and a cost relationship measured with three 2026 LLMs may not hold for their successors. No design choice removes this threat; we pin and publish every LLM identifier, revision, and serving parameter, frame the study as a snapshot of current-generation LLMs, and keep the instrument reproducible, so that replication with later LLMs requires only re-running a frozen pipeline.

\subsection{Construct Validity}
\label{sec:threats-construct}

\paragraph{Mono-operation bias on code quality.} CodeHealth is one proprietary composite, and using it alone risks measuring what that tool quantifies rather than code quality as such. The metric has been validated against organizational outcomes~\citep{tornhill_code_2022, borg_increasing_2024, borg_owns_2023} and benchmarked against human maintainability assessments~\citep{borg_ghost_2024}. Every one of these validation studies, and every study connecting CH to AI-assisted development that we cite~\citep{borg_echoes_2025, borg_code_2026, wirdemann_code_2026}, includes authors affiliated with CodeScene, the vendor of the metric, as does the present study. Our countermeasure was temporal: the metric, its tier boundaries, and the tool version were fixed when the sample was drawn, before any review agent had run, and recorded in the registration. Evidence from outside CodeScene is thinner but points the same way, as \citet{ramirez_exploring_2025} related a different composite, the Maintainability Index, to a different outcome, issue resolution on SWE-bench Lite, and found correlations with agent success whose strength varied across LLMs. A replication using an independent maintainability instrument would strengthen the construct.

\paragraph{Measurement point of the predictor.} Baseline CodeHealth is measured at the PR's head commit, before any agent runs, so the score includes what the PR itself contributed and not only the untouched repository. Because the tool scores whole files, we quantified both contributions in an exploratory measurement made after the outcomes were known. The median PR wrote 4.2\% of the content its score is computed over (an upper bound, since an edited line counts as written), 0.9\% in the unhealthy tier against 13.0\% in the healthy one, a gradient that follows file size; 18 PRs consist only of new files. Rescoring the same files at each PR's base commit, for the 446 PRs scorable at both commits, gives head and base scores correlating at \(r = .995\), with 94.6\% of PRs keeping their tier. Both checks favor \emph{clean up your existing codebase} over \emph{submit healthier patches} as the operative interpretation for most PRs.

\paragraph{File size: confound or mediator.} Unhealthy files are much larger than healthy ones (\Section{sec:results-filesize}), and the registered model adjusts for the lines a PR changes, not for the size of the files around them. The agents do not ingest whole files: their prompt contains only the diff and the file-reading tools page through a file (\Section{sec:method-instrument}), so a large file costs additional reads only when the context a change needs lies further inside it. Size-related smells, such as Large Method, Low Cohesion, and Bumpy Road, are components of the score, so adjusting for file size risks the overcontrol we avoid for footprint; \citet{ramirez_exploring_2025} similarly found a composite maintainability index tracking agent issue resolution more closely than its component metrics, SLOC among them. A size-adjusted refit absorbed the association for all three LLMs (\Section{sec:results-filesize}), which we interpret as mediation by file size (\Section{sec:discussion-filesize}). The residual threat is that the design cannot tell long, low-cohesion files apart from files that are long for other reasons, for which the mediation interpretation does not hold.

\paragraph{Mono-operation bias on cost.} Total tokens is one operationalization of a construct that also includes monetary spend, wall-clock latency, and downstream human attention, and these do not move together: cached input tokens are billed at a fraction of uncached ones, and a locally served LLM costs wall-clock time, hardware, and energy, not dollars per token. We pre-specify cache-neutral secondary outcomes and report the token composition alongside the total, but the token total is not a complete account of what agentic review costs an organization.

\paragraph{Cross-LLM non-comparability.} The three LLMs' prefix-cache regimes differ substantially, making token totals incomparable across LLMs (\Section{sec:method-models}). No cross-LLM inferential comparison is registered and the three fits are never pooled, so the threat bears on comparing magnitudes across LLMs, not on the within-LLM slopes that H1 tests.

\paragraph{Metric gaming.} If the agents could observe the metric they are measured against, any relationship between the two would be suspect. No review or fix agent receives the baseline CodeHealth score, the tier label, or any smell annotation, so the predictor cannot influence the instrument that produces the outcome. An earlier design that instructed the fix agent to raise CH to a target was dropped before registration for the reasons given in \Section{sec:experiment-design}; keeping the metric out of the instrument also removes this threat.

\paragraph{Prior LLM exposure to the code under review.} Every PR was merged between 2024-12-27 and 2025-07-30, before the one published training cutoff (January 2026, for \texttt{sonnet-5}), so the agents likely encountered sampled repositories, and possibly the merged patches, during pretraining. Contamination of this kind is documented in benchmarks built the same way~\citep{aleithan_swebench_2024}. Recognized code could cost fewer tokens to re-derive, which would bias \(\beta_{\mathit{health}}\) toward zero only if memorization is uncorrelated with CodeHealth, which we cannot establish, and contamination effects in code language models are not uniformly attenuating~\citep{cao_countermeasures_2024}. No agent sees the static analyzer's output, so contamination can reach the predictor only through a code property that drives both memorization and CH, and any such property at the repository level is absorbed by the random intercept. Agreement across three separately served LLMs is weak evidence against a shared artifact, since their training corpora overlap, and a membership probe remains the way to settle it.

\paragraph{Mono-method bias.} Every measurement in this study is automated. We do not assess whether the review comments are correct, whether the applied fixes are good, or whether a human reviewer would agree with either, so the study prices an activity whose value it does not establish. Assessing the correctness and value of the comments is outside the scope of this study: the pipeline stands in for the multi-agent review setups now in use, and is the instrument through which cost is measured, not the object of evaluation. Grading the comments is the first question we propose for future work (\Section{sec:discussion-future}).

\paragraph{Experimenter expectancy.} All agent prompts were designed by the authors, and prompt phrasing could in principle advantage some part of the design. The structural protection is that there are no conditions: the same prompts, tools, and loop run for every PR, CodeHealth tier, and LLM, so the prompts cannot favor one CH tier or LLM over another. The prompts were additionally hashed and published in full in the registration, filed before any review record had been read and before any fix session had run (\Section{sec:method-prereg}).

\subsection{Conclusion Validity}
\label{sec:threats-conclusion}

\paragraph{Low statistical power.} This is the study's principal conclusion threat. Every LLM's required sample size, extrapolated from its own pilot point estimate, exceeds the achieved 465 PRs, and the minimum detectable effect lies above the smallest pilot estimate (\Table{tab:pilot}), which is why H1 is a direction-consistency criterion across three instruments and not a significance test (\Section{sec:method-prereg}). All three intervals excluded zero (\Section{sec:results-rq1}), which the power analysis could not guarantee, and a true effect smaller than the achieved estimates would still have been missed.

\paragraph{Violated assumptions of statistical tests.} Token, comment, and iteration distributions are heavy-tailed; in pilot data a single PR accounted for approximately 45\% of a stratum's mean token total. We log-transform the cost and footprint variables in the mixed model, summarize distributions with medians and box plots instead of means, and use rank-based statistics for the smell correlations.

\paragraph{Fishing and error rate inflation.} With four research questions, three LLMs, five languages, and 19 smell categories observed in the sample, uncorrected exploration would almost certainly produce something significant. The registration fixed the confirmatory set before any review record had been read, the smell family is Holm-corrected, and everything labeled exploratory is controlled under Benjamini-Hochberg. We regard the registration as the main protection, in line with arguments about replication and researcher degrees of freedom in empirical computer science~\citep{cockburn_threats_2020}.

\paragraph{Non-independent observations.} Pooling the three LLMs' rows would triple the apparent sample while adding far less information, because the same diff is intrinsically easy or hard to review whichever LLM reviews it. On pilot data, naive pooling returned an apparently significant CodeHealth effect (\(p = .044\)) that moved to \(p = .19\) under PR-clustered and \(p = .25\) under repository-clustered standard errors. The registered analysis never pools, and we report this trap explicitly so that a reader attempting a re-analysis encounters it before repeating it.

\paragraph{Reliability of treatment implementation.} LLM outputs are not deterministic even at temperature 0, and an agentic tool loop amplifies the variation, so one run per PR inflates the residual variance against which the CodeHealth effect is estimated. The replication subsample quantifies this variance (\Table{tab:lambda}). For \texttt{deepseek-v4-flash} and \texttt{sonnet-5} it is small, so the narrow margin of \texttt{sonnet-5}'s interval reflects a small effect estimate, not run-to-run instability. For \texttt{qwen3-6} it is not, and that LLM's reported standard error does not include this run-level noise, so its true uncertainty is plausibly larger than \Table{tab:primarymodel} reports, without the sign or magnitude being biased. The subsample is small at 27 PRs, so the variance estimate is itself imprecise.

\paragraph{Attrition.} If pipeline failures were concentrated among the lowest-CodeHealth PRs, the surviving sample would differ systematically from the drawn one. Guard-stopped review sessions are retained as cost, not filtered (\Section{sec:method-attrition}), and for \texttt{qwen3-6} they are not random: they concentrate in lower-CH, higher-cost PRs, and that LLM's own interval excludes zero only with them retained (\Section{sec:results-sensitivity}). H1 does not depend on them, since the two other intervals still exclude zero; the residual threat is that \texttt{qwen3-6}'s individual result rests on the registered decision to count a spiraling review as cost.

\section{Conclusions}
\label{sec:conclusions}

A quality metric calibrated against human maintainability also predicts what a multi-agent LLM pipeline spends on reviewing and fixing code. In a pre-registered observational study of 465 agent-authored PRs from 246 repositories, processed by a frozen review-and-fix pipeline with three separately served LLMs, the CodeHealth coefficient was negative with a 95\% confidence interval excluding zero for all three LLMs, satisfying the registered criterion for \RQ{1}: one CH point was associated with 3.5\% to 6.1\% fewer tokens at constant footprint, language, and repository. Concerning \RQ{2}, the association is direct with respect to PR footprint, since unhealthy code did not arrive in larger PRs. Concerning \RQ{3}, Complex Method and Complex Conditional predicted fix-stage cost for every LLM, while Code Duplication, contrary to its registered prediction, did not. Concerning \RQ{4}, a single batch fix pass left median CH unchanged for every LLM and tier.

The exploratory analyses locate the association in the review stage. On unhealthy code the review agents take more turns and retrieve more context without raising more comments, so that review cost per finding is an estimated 30\% to 42\% lower for a typical healthy-tier PR than for an equivalently sized unhealthy one. The number of comments follows the count of code smells in the files under review, not the CodeHealth score. Adjusting for the size of the scored files absorbs the association for all three LLMs; CH and file size are too entangled in this sample to separate, and we interpret the association as mediated by file size. In absolute terms the effect is modest: the contrast between a typical unhealthy-tier and a typical healthy-tier PR corresponds to about 91,000 to 360,000 tokens for a single review-and-fix pass, depending on the LLM.

Of the two outcomes the study was designed to distinguish (\Section{sec:introduction}), the data supports the first, in a qualified form. Cost was not carried by PR footprint, so keeping PRs small does not substitute for code quality. The case for refactoring extends to agent-reviewed code, but the refactoring it supports is the kind that changes the structure an agent reads through, splitting long methods and low-cohesion modules into smaller units, and not changes that raise the score while leaving that structure in place (\Section{sec:discussion-roads-to-health}).

These conclusions are bounded by one quality metric, one cost construct, one pipeline, three LLMs at one point in their development, and merged agent-authored PRs in popular open-source repositories (\Section{sec:threats}). The published pipeline and replication package allow the association to be re-tested on later LLMs and with other code quality or maintainability metrics.

\backmatter

\bmhead{Funding}
This work was supported by the Swedish strategic research environment ELLIIT and by the Wallenberg AI, Autonomous Systems and Software Program (WASP) funded by the Knut and Alice Wallenberg Foundation. This work was partly funded by the NextG2Com Competence Centre – Next-Generation Communication and Computing Infrastructures and Applications – under the Vinnova grant number 2023-00541.

\bmhead{Acknowledgements}

We express our gratitude to CodeScene AB for providing access to CodeHealth MCP services for the duration of this research study.

\bmhead{Data Availability}

Source code and data are available at: \url{https://doi.org/10.5281/zenodo.22127920}.

\bmhead{Author Contributions}

Lo Gullstrand Heander: Conceptualization, Data curation, Formal analysis, Funding acquisition, Investigation, Methodology, Project administration, Software, Validation, Visualization, Writing -- original draft. Markus Borg: Resources, Validation, Writing -- review \& editing. Emma S\"oderberg: Funding acquisition, Project administration, Resources, Supervision, Writing -- review \& editing. Christofer Rydenf\"alt: Supervision, Writing -- review \& editing.

\bmhead{Conflict of Interest}
Markus Borg is employed by CodeScene, the company behind the CodeScene analysis tool used in this study, including the CodeHealth metric. His contributions to this work were resources, validation, and reviewing and editing the manuscript (see Author Contributions). The authors declare that no commercial interests compromised the scientific rigor of the study design, data collection, or analysis.

\bmhead{Ethics Approval and Consent}
Not applicable; the study involves no human participants and no personal data.

\bmhead{Clinical Trial Number}
Clinical trial number: not applicable.

\bibliography{refs}%

\end{document}